\documentclass[12pt]{article}
\usepackage{graphicx}
\usepackage{latexsym,amsmath,amsfonts,amssymb}
\newcommand{\be}{\begin{equation}}
\newcommand{\ee}{\end{equation}}
\newcommand{\beq}{\begin{equation}}
\newcommand{\eeq}{\end{equation}}
\newcommand{\bea}{\begin{eqnarray}}
\newcommand{\eea}{\end{eqnarray}}

\newcommand{\ba}{\begin{eqnarray}}
\newcommand{\ea}{\end{eqnarray}}
\begin{document}
\baselineskip=15.5pt
\pagestyle{plain}
\setcounter{page}{1}


\def\del{{\partial}}
\def\vev#1{\left\langle #1 \right\rangle}
\def\cn{{\cal N}}
\def\co{{\cal O}}
\def\IC{{\mathbb C}}
\def\IR{{\mathbb R}}
\def\IZ{{\mathbb Z}}
\def\RP{{\bf RP}}
\def\CP{{\bf CP}}
\def\Poincare{{Poincar\'e }}
\def\tr{{\rm tr}}
\def\tp{{\tilde \Phi}}

\newcommand{\bA}{{\bf A }}
\newcommand{\bN}{{\bf N }}

\def\TL{\hfil $\displaystyle{##}$ }
\def\TR{$\displaystyle{{}##}$ \hfil}
\def\TC{\hfil $\displaystyle{##}$\hfil}
\def\TT{\hbox{##}}
\def\HLINE{\noalign{\vskip1\jot}\hline\noalign{\vskip1\jot}}
\def\seqalign#1#2{\vcenter{\openup1\jot
  \halign{\strut #1\cr #2 \cr}}}
\def\lbldef#1#2{\expandafter\gdef\csname #1\endcsname {#2}}
\def\eqn#1#2{\lbldef{#1}{(\ref{#1})}%
\begin{equation} #2 \label{#1} \end{equation}}
\def\eqalign#1{\vcenter{\openup1\jot
    \halign{\strut\span\TL & \span\TR\cr #1 \cr
   }}}
\def\eno#1{(\ref{#1})}
\def\href#1#2{#2}
\def\half{{1 \over 2}}



\def\NO{\nonumber}

\def\bea{\begin{eqnarray}}
\def\eea{\end{eqnarray}}

\def\beqx{\begin{displaymath}}
\def\eeqx{\end{displaymath}}

\newcommand{\bmat}{\left(\begin{array}}
\newcommand{\emat}{\end{array}\right)}

\def\half{\frac{1}{2}}


\newtheorem{definition}{Definition}[section]
\newtheorem{theorem}{Theorem}[section]
\newtheorem{lemma}{Lemma}[section]
\newtheorem{corollary}{Corollary}[section]


\def\a{\alpha}
\def\b{\beta}
\def\c{\chi}
\def\d{\delta}
\def\e{\epsilon}
\def\f{\phi}
\def\g{\gamma}
\def\h{\eta}
\def\i{\iota}
\def\j{\psi}
\def\k{\kappa}
\def\l{\lambda}
\def\m{\mu}
\def\n{\nu}
\def\o{\omega}
    \def\om{\omega}
\def\p{\pi}
\def\q{\theta}
\def\th{\theta}
\def\r{\rho}
\def\s{\sigma}
\def\t{\tau}
\def\x{\xi}
\def\z{\zeta}
\def\D{\Delta}
\def\F{\Phi}
\def\G{\Gamma}
\def\J{\Psi}
\def\L{\Lambda}
\def\O{\Omega}
    \def\Om{\Omega}
\def\P{\Pi}
\def\Q{\Theta}
    \def\Th{\Theta}
\def\S{\Sigma}
\def\U{\Upsilon}
\def\X{\Xi}


\def\ve{\varepsilon}
\def\vr{\varrho}
\def\vs{\varsigma}
\def\vq{\vartheta}
    \def\vth{\vartheta}
\def\tvf{\tilde{\varphi}}
\def\vf{\varphi}
    \def\vphi{\varphi}


\def\ca{{\cal A}}
\def\cb{{\cal B}}
\def\cc{{\cal C}}
\def\cd{{\cal D}}
\def\ce{{\cal E}}
\def\cf{{\cal F}}
\def\cg{{\cal G}}
\def\ch{{\cal H}}
\def\ci{{\cal I}}
\def\cj{{\cal J}}
\def\ck{{\cal K}}
\def\cl{{\cal L}}
\def\cm{{\cal M}}
\def\cn{{\cal N}}
\def\co{{\cal O}}
\def\cp{{\cal P}}
\def\cq{{\cal Q}}
\def\car{{\cal R}}
\def\cs{{\cal S}}
\def\ct{{\cal T}}
\def\cu{{\cal U}}
\def\cv{{\cal V}}
\def\cw{{\cal W}}
\def\cx{{\cal X}}
\def\cy{{\cal Y}}
\def\cz{{\cal Z}}



\def\Sc#1{{\hbox{\sc #1}}}      
\def\Sf#1{{\hbox{\sf #1}}}      
\def\mb#1{\mbox{\boldmath $#1$}}


\def\slpa{\slash{\pa}}                         
\def\slin{\SLLash{\in}}                                 
\def\bo{{\raise-.3ex\hbox{\large$\Box$}}}               
\def\cbo{\Sc [}                                         
\def\pa{\partial}                                       
\def\de{\nabla}                                         
\def\dell{\nabla}                                       
\def\su{\sum}                                           
\def\pr{\prod}                                          
\def\iff{\leftrightarrow}                               
\def\conj{{\hbox{\large *}}}                            
\def\ltap{\raisebox{-.4ex}{\rlap{$\sim$}} \raisebox{.4ex}{$<$}}   
\def\gtap{\raisebox{-.4ex}{\rlap{$\sim$}} \raisebox{.4ex}{$>$}}   
\def\face{{\raise.2ex\hbox{$\displaystyle \bigodot$}\mskip-2.2mu \llap
{$\ddot
        \smile$}}}                                   
\def\dg{\dagger}                                     
\def\ddg{\ddagger}                                   
\def\trans{\mbox{\scri T}}                           
\def\>{\rangle}                                      
\def\<{\langle}                                      


\def\sp#1{{}^{#1}}                                   
\def\sb#1{{}_{#1}}                                   
\newcommand{\sub}[1]{\phantom{}_{(#1)}\phantom{}}    
\newcommand{\supt}[1]{\phantom{}^{(#1)}\phantom{}}    
\def\oldsl#1{\rlap/#1}                               
\def\slash#1{\rlap{\hbox{$\mskip 1 mu /$}}#1}        
\def\Slash#1{\rlap{\hbox{$\mskip 3 mu /$}}#1}        
\def\SLash#1{\rlap{\hbox{$\mskip 4.5 mu /$}}#1}      
\def\SLLash#1{\rlap{\hbox{$\mskip 6 mu /$}}#1}       
\def\wt#1{\widetilde{#1}}                            
\def\Hat#1{\widehat{#1}}                             
\def\lbar#1{\ensuremath{\overline{#1}}}              
\def\VEV#1{\left\langle #1\right\rangle}             
\def\abs#1{\left| #1\right|}                         
\def\leftrightarrowfill{$\mathsurround=0pt \mathord\leftarrow \mkern-6mu
        \cleaders\hbox{$\mkern-2mu \mathord- \mkern-2mu$}\hfill
        \mkern-6mu \mathord\rightarrow$}        
\def\dvec#1{\vbox{\ialign{##\crcr
        $\hfil\displaystyle{#1}\hfil$\crcr}}}           
\def\dt#1{{\buildrel {\hbox{\LARGE .}} \over {#1}}}     
\def\dtt#1{{\buildrel \bullet \over {#1}}}              
\def\der#1{{\pa \over \pa {#1}}}                        
\def\fder#1{{\d \over \d {#1}}}                         
\def\tr{{\rm tr \,}}                                    
\def\Tr{{\rm Tr \,}}                                    
\def\diag{{\rm diag \,}}                                
\def\preal{{\rm Re\,}}                                  
\def\pim{{\rm Im\,}}                                    


\def\partder#1#2{{\partial #1\over\partial #2}}        
\def\parvar#1#2{{\d #1\over \d #2}}                    
\def\secder#1#2#3{{\partial^2 #1\over\partial #2 \partial #3}}  
\def\on#1#2{\mathop{\null#2}\limits^{#1}}              
\def\bvec#1{\on\leftarrow{#1}}                         
\def\oover#1{\on\circ{#1}}                             


\def\Deq#1{\mbox{$D$=#1}}                               
\def\Neq#1{\mbox{$cn$=#1}}                              
\newcommand{\ampl}[2]{{\cal M}\left( #1 \to #2 \right)} 


\def\NPB#1#2#3{Nucl. Phys. B {\bf #1} (19#2) #3}
\def\PLB#1#2#3{Phys. Lett. B {\bf #1} (19#2) #3}
\def\PLBold#1#2#3{Phys. Lett. {\bf #1}B (19#2) #3}
\def\PRD#1#2#3{Phys. Rev. D {\bf #1} (19#2) #3}
\def\PRL#1#2#3{Phys. Rev. Lett. {\bf #1} (19#2) #3}
\def\PRT#1#2#3{Phys. Rep. {\bf #1} C (19#2) #3}
\def\MODA#1#2#3{Mod. Phys. Lett.  {\bf #1} (19#2) #3}


\def\norder{\raisebox{-.13cm}{\ensuremath{\circ}}\hspace{-.174cm}\raisebox{.13cm}{\ensuremath{\circ}}}
\def\bz{\bar{z}}
\def\bw{\bar{w}}
\def\-{\hphantom{-}}
\newcommand{\dd}{\mbox{d}}
\newcommand{\scr}{\scriptscriptstyle}
\newcommand{\scri}{\scriptsize}
\def\rand#1{\marginpar{\tiny #1}}               
\newcommand{\rstar}{\rand{\bf\large *}}
\newcommand{\rup}{\rand{$\uparrow$}}
\newcommand{\rdown}{\rand{$\downarrow$}}


\def\ads{{\it AdS}}
\def\adsp{{\it AdS}$_{p+2}$}
\def\cft{{\it CFT}}

\newcommand{\ber}{\begin{eqnarray}}
\newcommand{\eer}{\end{eqnarray}}

\newcommand{\beqar}{\begin{eqnarray}}
\newcommand{\cN}{{\cal N}}
\newcommand{\cO}{{\cal O}}
\newcommand{\cA}{{\cal A}}
\newcommand{\cT}{{\cal T}}
\newcommand{\cF}{{\cal F}}
\newcommand{\cC}{{\cal C}}
\newcommand{\cR}{{\cal R}}
\newcommand{\cW}{{\cal W}}
\newcommand{\eeqar}{\end{eqnarray}}
\newcommand{\tht}{\thteta}
\newcommand{\lm}{\lambda}\newcommand{\Lm}{\Lambda}
\newcommand{\eps}{\epsilon}


\newcommand{\nonu}{\nonumber}
\newcommand{\oh}{\displaystyle{\frac{1}{2}}}
\newcommand{\dsl} {\kern.06em\hbox{\raise.15ex\hbox{$/$}\kern-.56em\hbox{$\partial$}}}
\newcommand{\id}{i\!\!\not\!\partial}
\newcommand{\as}{\not\!\! A}
\newcommand{\ps}{\not\! p}
\newcommand{\ks}{\not\! k}
\newcommand{\dv}{d^2x}
\newcommand{\Z}{{\cal Z}}
\newcommand{\N}{{\cal N}}
\newcommand{\Dsl}{\not\!\! D}
\newcommand{\Bsl}{\not\!\! B}
\newcommand{\Psl}{\not\!\! P}
\newcommand{\eeqarr}{\end{eqnarray}}
\newcommand{\ZZ}{{\rm \kern 0.275em Z \kern -0.92em Z}\;}


\def\del{{\delta^{\hbox{\sevenrm B}}}} \def\ex{{\hbox{\rm e}}}
\def\azb{A_{\bar z}} \def\az{A_z} \def\bzb{B_{\bar z}} \def\bz{B_z}
\def\czb{C_{\bar z}} \def\cz{C_z} \def\dzb{D_{\bar z}} \def\dz{D_z}
\def\im{{\hbox{\rm Im}}} \def\mod{{\hbox{\rm mod}}} \def\tr{{\hbox{\rm
Tr}}}
\def\ch{{\hbox{\rm ch}}} \def\imp{{\hbox{\sevenrm Im}}}
\def\trp{{\hbox{\sevenrm Tr}}} \def\vol{{\hbox{\rm Vol}}}
\def\rl{\Lambda_{\hbox{\sevenrm R}}} \def\wl{\Lambda_{\hbox{\sevenrm W}}}
\def\fc{{\cal F}_{k+\cox}} \def\vev{vacuum expectation value}
\def\nodiv{\mid{\hbox{\hskip-7.8pt/}}}
\def\ie{{\em i.e.}}
\def\ie{\hbox{\it i.e.}}

\def\CC{{\mathchoice
{\rm C\mkern-8mu\vrule height1.45ex depth-.05ex
width.05em\mkern9mu\kern-.05em}
{\rm C\mkern-8mu\vrule height1.45ex depth-.05ex
width.05em\mkern9mu\kern-.05em}
{\rm C\mkern-8mu\vrule height1ex depth-.07ex
width.035em\mkern9mu\kern-.035em}
{\rm C\mkern-8mu\vrule height.65ex depth-.1ex
width.025em\mkern8mu\kern-.025em}}}

\def\RR{{\rm I\kern-1.6pt {\rm R}}}
\def\NN{{\rm I\!N}}
\def\ZZ{{\rm Z}\kern-3.8pt {\rm Z} \kern2pt}
\def\IB{\relax{\rm I\kern-.18em B}}
\def\ID{\relax{\rm I\kern-.18em D}}
\def\II{\relax{\rm I\kern-.18em I}}
\def\IP{\relax{\rm I\kern-.18em P}}
\newcommand{\CS}{{\scriptstyle {\rm CS}}}
\newcommand{\CSs}{{\scriptscriptstyle {\rm CS}}}
\newcommand{\rc}{\nonumber\\}
\newcommand{\bear}{\begin{eqnarray}}
\newcommand{\eear}{\end{eqnarray}}
\newcommand{\W}{{\cal W}}
\newcommand{\LL}{{\cal L}}

\def\mani{{\cal M}}
\def\calo{{\cal O}}
\def\calb{{\cal B}}
\def\calw{{\cal W}}
\def\calz{{\cal Z}}
\def\cald{{\cal D}}
\def\calc{{\cal C}}
\def\to{\rightarrow}
\def\ele{{\hbox{\sevenrm L}}}
\def\ere{{\hbox{\sevenrm R}}}
\def\zb{{\bar z}}
\def\wb{{\bar w}}
\def\nodiv{\mid{\hbox{\hskip-7.8pt/}}}
\def\menos{\hbox{\hskip-2.9pt}}
\def\dr{\dot R_}
\def\drr{\dot r_}
\def\ds{\dot s_}
\def\da{\dot A_}
\def\dga{\dot \gamma_}
\def\ga{\gamma_}
\def\dal{\dot\alpha_}
\def\al{\alpha_}
\def\cls{{closing}}
\def\vev{vacuum expectation value}
\def\tr{{\rm Tr}}
\def\to{\rightarrow}
\def\too{\longrightarrow}


\def\a{\alpha}
\def\b{\beta}
\def\c{\gamma}
\def\d{\delta}
\def\e{\epsilon}           
\def\f{\phi}               
\def\vf{\varphi}  \def\tvf{\tilde{\varphi}}
\def\vp{\varphi}
\def\g{\gamma}
\def\h{\eta}
\def\i{\iota}
\def\j{\psi}
\def\k{\kappa}                    
\def\l{\lambda}
\def\m{\mu}
\def\n{\nu}
\def\o{\omega}  \def\w{\omega}
\def\q{\theta}  \def\th{\theta}                  
\def\r{\rho}                                     
\def\s{\sigma}                                   
\def\t{\tau}
\def\u{\upsilon}
\def\x{\xi}
\def\z{\zeta}
\def\pt{\tilde{\varphi}}
\def\tt{\tilde{\theta}}
\def\lab{\label}
\def\6{\partial}
\def\wg{\wedge}
\def\atanh{{\rm arctanh}}
\def\bpsi{\bar{\psi}}
\def\bt{\bar{\theta}}
\def\bvf{\bar{\varphi}}

%

\newfont{\namefont}{cmr10}
\newfont{\addfont}{cmti7 scaled 1440}
\newfont{\boldmathfont}{cmbx10}
\newfont{\headfontb}{cmbx10 scaled 1728}
\renewcommand{\theequation}{{\rm\thesection.\arabic{equation}}}
\begin{titlepage}

\rightline{MCTP-11-01}
\rightline{UTTG-01-11}
\begin{center} \Large \bf  Heating up the Baryonic Branch with U-duality:\\
a unified picture of conifold black holes

\end{center}

\vskip 0.15truein
\begin{center}
Elena C\'aceres${}^{*}$\footnote{elenac@zippy.ph.utexas.edu},
Carlos
N\'u\~nez${}^{\dagger}$\footnote{c.nunez@swansea.ac.uk} and Leopoldo A.
Pando Zayas
${}^{**}$\footnote{lpandoz@umich.edu}
\vspace{0.4in}\\
${}^{*}$ \it{Facultad de Ciencias \\
Universidad de  Colima\\
Bernal Diaz del Castillo 340, Colima, M\'exico.\\
\small{and}\\
Theory Group, Department of Physics,\\
University of Texas at Austin,
Austin, TX 78727, USA.\\
}
\vspace{0.2in}
${}^{\dagger}$ \it{Department of Physics\\ University of Swansea, Singleton
Park\\
Swansea SA2 8PP
\\ United Kingdom.\\}
\vspace{0.2in}
${}^{**}$ \it{Michigan Center
for Theoretical Physics\\Randall Laboratory of Physics,
the University of Michigan\\
Ann Arbor, MI 48109-1040. USA}
\vspace{0.2in}
\end{center}
\centerline{{\bf Abstract}}
We study different aspects of a U-duality  recently
presented by Maldacena and Martelli
and apply it to  non-extremal backgrounds. In particular,
starting from new non-extremal
wrapped D5 branes we generate  new non-extremal  generalizations
of the Baryonic Branch
of the Klebanov-Strassler solution.
We also elaborate on different conceptual aspects of these
U-dualities,
like its action on (extremal and non-extremal) Dp branes,
dual models
for Yang-Mills-like theories, generic asymptotics and
decoupling limit
of the generated  solutions.

\vskip1truecm
\vspace{0.1in}
\smallskip
\end{titlepage}
\setcounter{footnote}{0}
\tableofcontents

\section{Introduction}
The Maldacena conjecture \cite{Maldacena:1997re} and some of its
refinements \cite{Gubser:1998bc} are the guiding principle behind much
of the progress in the interface String Theory-Quantum Field theory in
the last twelve years. The influence of this approach  extends to
toy models with different number of supersymmetries (SUSY's), systems at finite temperature
and/or finite density and lower dimensional systems. The applications to
physically relevant systems do
not seem to be exhausted and on the contrary, increase with time. In this
sense, finding new
(trustable) solutions
to the equations of motion of  String Theory (even in the point
particle and classical approximation) has become
an industry with various applications in Physics and Mathematics. Solution
generating
techniques have certainly played a role. Some examples worth mentioning
are the
combination of T-dualities  and shifts of coordinates that generated
solutions dual to minimally SUSY superconformal field theories (beta
deformations) or duals to non-relativistic field theories
\cite{Lunin:2005jy}\footnote{In other areas of Physics, solution
generating techniques play a major role. For example, many of the
interesting solutions in higher dimensional gravity were constructed
using solution generating algorithms. See for example
\cite{Belinski:2001ph}
for original papers  and \cite{Elvang:2007rd}
for a nice summary with an important application.}.

In this paper we will focus our attention on a particular solution
generating technique that was presented in
\cite{Maldacena:2009mw}. The procedure
suggested by these authors
starts by taking
a solution in type IIB string theory proposed to be dual to
(a suitably UV-completed
version of) $\cn =1$ Super Yang-Mills in four dimensions. The solution
has the topology $R^{1,3}\times M_6$ where $M_6$ preserves four
supercharges. The algorithm to generate the new solutions goes as follows:
first apply a set of three
T-dualities in the $R^{3}$ directions, which leaves us with a IIA
configuration; lift this configuration to M-theory and boost--with rapidity
$\beta$--in the
eleventh direction; then reduce to IIA and T-dualize back in $R^3$. This
generates a solution that roughly speaking is the dual description to
the baryonic branch \cite{Butti:2004pk}
of the Klebanov-Strassler field theory \cite{Klebanov:2000hb}. For all the
technical details of this procedure the reader can refer to
\cite{Maldacena:2009mw}
or to our Appendix \ref{appendix1}.

The `seed solution' (as we will refer to the initial solution on
which the generating algorithm is applied) was discussed in the papers
\cite{Casero:2006pt} (Section 8) and \cite{HoyosBadajoz:2008fw}
(Section 4.3). In a particular limit, this  becomes the exact solution
discussed in \cite{Chamseddine:1997nm}.

We will also refer to the
solution generating algorithm, that is a U-duality,
as `rotation'  in a sense explained below and in previous papers.
In the following we will emphasize some aspects of the generating
technique presented in \cite{Maldacena:2009mw}
that we find particularly interesting and have not been explicitly
discussed in the existing bibliography:

\begin{itemize}
\item{In the case of \cite{Maldacena:2009mw},
the rotation generates D3 brane charge. This is the reason why the
$SO(1,3)$ isometry of the background is untouched, in spite of doing
{\it different} operations in time (boost) and in the $R^{3}$ directions
(T-dualities). As a by-product of the generation of D3 branes the dilaton
is invariant under this set of operations.}

\item{It can be seen that this U-duality (sequence of T-dualities, lifts
and
boosts) described above is equivalent to a {\it particular case} of
`rotation' in
the $SU(3)$ structure of the
manifold $M_6$ which is characterized by a complex 3-form $\Omega_3$ and a 2-form $J_2$ - see the papers
\cite{Minasian:2009rn} and  \cite{Gaillard:2010qg}. This is another way of
understanding the presence of the $SO(1,3)$ isometry: from this
perspective, all the rotation
`occurs' in the internal $M_6$.
}

\item{The rotation generates a solution that contains two free
parameters, the
value of the dilaton at infinity (already present in the seed solution)
and the
boost parameter $\beta$. In order for the final background to be dual to
the Klebanov-Strassler QFT {\it decoupled } from gravity, we must take
the limit $\beta\to\infty$ so that the generated warp factor vanishes
asymptotically. In this sense, the seed solution describes a field theory
coupled to gravity. Only in a particular limit-see the discussion in
\cite{Maldacena:2009mw}, \cite{Gaillard:2010qg},
we approach the near brane
solution of \cite{Chamseddine:1997nm}.
}

\item{This rotation or U-duality, is a very curious operation from the
viewpoint of the dual
field theory. Indeed, it generates (aside from the bifundamental fields) new
global symmetries, like the baryonic $U(1)$.}

\item{In the same vein: the connection between the two field
theories from a  geometric viewpoint (Field theory at strong coupling!)
is just the mentioned U-duality starting
from
the $D5$-brane side. From a (weakly interacting) field theoretic
viewpoint the operation is very subtle \cite{Maldacena:2009mw}. Indeed, it
involves a second order
fluctuation expansion (mass spectrum)
in the classical and weakly coupled F and D term equations on the quiver
side \cite{Dymarsky:2005xt}, that is matched
with a classical and weakly coupled expansion of the twisted
compactification of the theory in the wrapped D5 brane
mass spectrum \cite{Andrews:2006aw}.
}
\item{After the first set of T-dualities described above, we have a IIA
background with D2 brane charges. When lifted to M-theory we
have M2 branes, that after boosted  generate M2 charge and an M-wave.
When reduced, this generates a
D0, D2 and NS $H_3$ field. The generation of the $H_3$ in the presence of RR
three-form implies the need to generate
$F_5$. See Appendix \ref{appendix1} for technical details.}

\item{We emphasized that the U-duality ( chain of T-dualities,
lifts
and boost) described above is a {\it particular case} of the rotation of
$J_2,
\Omega_3$ discussed in \cite{Minasian:2009rn},
\cite{Gaillard:2010qg}, \cite{Halmagyi:2010st}. On the other hand, the
rotation of $J_2,\Omega_3$ relies on the background being supersymmetric,
while the
U-duality can be also applied to Non-SUSY backgrounds. This
will play an important role in the rest of this paper.}

\end{itemize}
\subsection{General Idea of this Paper}
In this paper we will try to gain a different perspective on this solution
generating technique. We will not focus on the rotation of $J,\Omega$ but
rely on the U-duality description (in spite of the latter being a particular
case of the former). Our interests will be two-fold. On one hand, we will
try to get a better handle on the generating algorithm by changing it
(suggesting other related algorithms) and
applying it to various cases. On the other hand, we will apply it to
backgrounds that do not preserve SUSY. We will study the effect of the
rotation on black hole solutions, generating {\it new} non-extremal
solutions with
horizons and other curious features that will be discussed.
We will also start the study of the properties of those
newly generated
solutions.

This paper is organized as follows: In sections
\ref{warmupz} and \ref{psection} as a warm-up
exercise we discuss a chain of dualities and boost
acting on non-extremal Dp branes.
In section \ref{wittensakaisugimotosection}
we will apply this to rotate a background dual
to a version of Yang-Mills. Then, in section \ref{pminusqsection}
 we will study a variation
of this U-duality that will clarify various aspects of the coming
material. In Section \ref{rotatingn=1} we will U-dualize
a new solution describing the non-extremal deformation
of a stack of $N_c$ D5 branes
wrapping a two-cycle inside the resolved conifold.
We will elaborate upon
various aspects of this particular rotation in section
\ref{analysisinteresting}.
Finally in Section \ref{seccion4} we comment on
decoupling aspects of the rotation.
Various appendices complement our presentation. They have been
written with plenty of detail hoping that colleagues
working on these topics will find them useful. We close
with a summary, conclusions and a list of possible future projects.

\section{A warm-up example: Rotation of Dp-branes}\label{warmupz}

In this section, we will start with a simple example
of the `rotation' procedure.
That is, a sequence of T-dualities, bringing the background to a type IIA
solution of the supergravity equations of motion, followed by a lift to
eleven dimensions, where we will boost the
configuration. We will then reduce to IIA and T-dualize back, to
what we will call the `rotated background'. As a toy example, in this
section we will rotate
flat $Dp$ branes; first in
$p$ directions and then in $p-q$ directions. We will do this in detail to
appreciate the differences this introduces in the generated solution.
In Appendix \ref{anothersolgen}, we will propose and analyze another
possible U-duality to
generate new
solutions.

\subsection{``Rotation'' in p directions}\label{psection}

Consider backgrounds of IIA/IIB of the form,
\bea
& & ds^2= H(\r)^{-1/2}\Big[ - h(\r)dt^2 + d\vec{x}_p^2  \Big] +
 H(\r)^{1/2}\Big[\frac{d\r^2}{h(\r)}+ \r^2 d\Omega_{8-p}   \Big],
\nonumber\\
& & e^{2\phi[initial]}= e^{2\phi(\infty)}H^{\frac{3-p}{2}},\nonumber\\
& & *F_{p+2}= \mathcal{ Q }\textit{Vol}_{\  \Omega_{8-p}}, \nonumber\\
& & F_{p+2}=-\partial_\r A(\r) dt\wedge dx_1\wedge....\wedge dx_p\wedge d\r
\label{flatdp}
\eea
with,
\beq
H(\r)=1+\Big(\frac{L_p}{\r}\Big)^{7-p},\;\; h(\r)= 1-\Big(\frac{R_T}{\r}\Big)^{7-p},\;\; A(\r)=
\frac{\alpha }{g_s H(\r)},
 \eeq
where $\alpha= \frac{\tilde Q}{ L_p ^{7-p}}$ and $\mathcal{Q} =(7-p) \tilde Q$ is related to the charge of the solution (see Appendix \ref{Dpblack} for details).
If we choose $R_T=0$, the configuration above is
typically singular (except for $p=3$) and preserves 16 SUSYs.
Recall that for even (odd) values of $p$, we are
dealing with a solution of Type IIA(B) supergravity.

Now, we will `rotate' this background,
by first T-dualizing in the $p$ directions, this will bring us to a  IIA
solution, we will lift the solution to
eleven-dimensional supergravity and perform a boost of rapidity $\beta$
in the eleventh-direction.
We will then reduce to IIA and T-dualize back
in the $\vec{x}_p$ coordinates to obtain what we call the
`rotated background'.
Our final
rotated background is given by\footnote{In Appendix \ref{facil},
we will present the
intermediate steps of this calculation for the extremal case $h=1$.
},
\bea
& & ds^2=\frac{ H(\r)^{-1/2} }{R(\r)}\Big[ -  h(\r)dt^2 + d\vec{x}_p^2  \Big] +
R(\r) H(\r)^{1/2}\Big[\frac{d\r^2}{h(\r)}+ \r^2 d\Omega_{8-p}   \Big]
\label{eq:rotmetric}\\
 & &\nonumber\\
& & e^{2\phi[final]}=
e^{2\phi[initial]}R(\r)^{3-p}=e^{2\phi(\infty)}(H(\r) R(\r)^2)^{\frac{3-p}{2}}\label{eq:rotdilaton}\\
& & F_{p+2}=-\partial_\r a(\r) dt\wedge dx_1\wedge....\wedge dx_p\wedge d\r
\label{flatdproated}
\eea
where we have defined,
\bea
& & R^2= (A(\r)\sinh\beta+\cosh\beta)^2 -\frac{h(\r) \sinh^2\beta}{g_s^2 H(\r)^2},\label{eq:r}\\
& & a=\frac{1}{R^2}[A(\r)\cosh 2\beta + (A(\r)^2+1)\cosh\beta\sinh\beta -\frac{h(\r)}{2 g_s^2 H(\r)^2}\sinh(2\beta)]
\label{zakama}
\eea
We have explicitly checked that the rotated background is a solution
of the equations  of motion. Notice that eq. (\ref{eq:r}) can be written as
\be \label{eq:rsimp}
R^2 =  \cosh^2\beta + A(\r) \sinh 2\beta +(\frac{R_T}{L_p})^{7-p} \frac{\sinh^2\beta}{H(\rho) g_s ^2}
\ee
which makes clear  that $R$ is strictly positive.
In a similar way, eq. (\ref{zakama}) can be written as
\bea\label{eq:a}
 a&=&\frac{1}{H R^2}\left[ \frac{1}{g_s} \alpha \cosh(2\, \beta)  +\left(\frac{R_T}{L_p}\right)^{7-p} \frac{\sinh\beta \cosh\beta}{g_s ^2}+ H(\r)\sinh\beta \cosh\beta\right].
\eea
The metric (\ref{eq:rotmetric})  has the structure of a warped space,
\be\label{eq:rotmetricnewH}
ds^2= \mathcal{H}(\r)^{-1/2} \Big[ -  h(\r)dt^2 + d\vec{x}_p^2  \Big] +
\mathcal{H}(\r)^{1/2}\Big[\frac{d\r^2}{h(\r)}+ \r^2 d\Omega_{8-p}   \Big],
\ee
where the new warp factor is
\bea
\mathcal{H}(\r)=H(\r) R(\r)^2=H(\r)\cosh^2 \beta + \frac{\alpha}{g_s}
\sinh 2\beta
+(\frac{R_T}{L_p})^{7-p} \frac{\sinh^2 \beta}{ g_s ^2}
\ea
Note that $\mathcal{H}(\r)$ is a harmonic function of the transverse space.
Hence, after the rotation in $p$ directions we are left with a $Dp$ brane
solution. Indeed, in terms of the new warp factor, the dilaton and gauge potential are
\begin{align}\label{eq:rotdila}
&e^{2\phi[final]}= e^{2\phi(\infty)}\mathcal{H}(\r)^{\frac{3-p}{2}}\nonumber\\
&\partial_\r a(\r) =\left [\frac{\alpha}{g_s}  +(\frac{R_T}{L_p})^{7-p}
\frac{\tanh\beta}{g_s^2}\right]\ \partial_\r
\Big(\frac{1}{\mathcal{H}(\r)}\Big) =\tilde\alpha\  \partial_\r\Big(\frac{1}{\mathcal{H}(\r)}\Big),
\end{align}
which together with (\ref{eq:rotmetricnewH}) define a $Dp$ brane background  with a $\beta$ dependent  RR charge.
At infinity the warp factor asymptotes to a constant,
\be
\mathcal{H}(\r) \sim \cosh^2 \beta + \frac{\alpha}{g_s}  \sinh 2\beta +(\frac{R_T}{L_p})^{7-p} \frac{\sinh^2\beta}{ g_s ^2},
\label{vvvvxv}\ee
and the space is asymptotically flat, as expected.
However, it is interesting to note that even if we start with the Dp branes after the decoupling limit is taken, that is
$H=\frac{L_p^{7-p}}{\r^{7-p}}$, we will have in the UV that,
\beq
\mathcal{H}(\r) \sim \frac{L_p^{7-p}}{\r^{7-p}} \cosh^2\beta + \frac{\alpha}{g_s}  \sinh 2\beta +(\frac{R_T}{L_p})^{7-p} \frac{\sinh^2\beta}{ g_s ^2}
\eeq
So, again, the warp factor asymptotes to a constant. In other words, this rotation is taking the
configuration \emph{ out of the decoupling limit} or coupling
the field theory modes to gravity.

As is well known, before the rotation, the charge of the Dp brane solution
is quantized.
After the rotation the charge is,
\begin{align}
\frac{1}{(2 \pi l_s)^{7-p} }\int_{S^{8-p}} *F_{p+2}&=  \tilde \alpha \  \r ^{8-p}\mathcal{H}'(\r) Vol_{S^{8-p}}\nonumber\\
& = \textrm{Q} \cosh^2 \beta\,   + {R_T}^{7-p} \frac{c_p}{2 g_s}\sinh 2\beta
\end{align}
where $c_p= \frac{(7-p)Vol_{S^{8-p}}}{(2 \pi l_s)^{7-p}g_s}$. Note that, generically,  the charge is not  quantized. This
is not unusual since the supergravity dualities
involved in the rotation
procedure are a symmetry of the supergravity
equations and not of the full string theory.

We will now move to study a
more interesting example from the viewpoint of gauge-strings duality.
We will apply the rotation
procedure to the a non-SUSY Yang-Mills-like theory,
first presented in \cite{Witten:1998zw} and further
studied in \cite{Sakai:2004cn}, \cite{Aharony:2006da}.
\subsubsection{Example: rotation of a dual to a Yang-Mills-like
theory}\label{wittensakaisugimotosection}
The original background, consists of the decoupling limit of a stack of
$N_c$ D4 branes wrapping a circle
 with SUSY breaking boundary conditions. It was
discussed with details in various
publication, see for example  \cite{Witten:1998zw}, \cite{Sakai:2004cn},
\cite{Aharony:2006da}.
We summarize it here (in string frame)\footnote{ In \cite{Sakai:2004cn} the authors performed a rescaling of the RR potential:
$C_{p +1} \rightarrow \frac{\kappa_0^2 \mu_{6-p}}{\pi} C_{p+1} $ and as  a result the RR charge in their paper is  measured in units of $2\pi$. We do not perform such rescaling here. }
\bea
 & & ds^2= H(\r)^{-1/2}\Big[ - h(\r)dt^2 + d\vec{x}_{123}^2 +f(\r)dx_4^2  \Big]+
 H(\r)^{1/2}\Big[\frac{d\r^2}{f(\r)h(\r)}+ \r^2 d\Omega_{4}   \Big],
\nonumber\\
& & e^{2\phi[initial]}= g_s^2 H^{-\frac{1}{2}},\nonumber\\
& & F_{6}=-\partial_\r A(\r) dt\wedge dx_1\wedge....\wedge dx_4\wedge d\r \ \ \ \  F_4= * F_6
 \label{D4witten}
 \eea
 where $A(\r)=\frac{1 }{ g_s H(\r) }$ and $Vol_4 $ is the volume of the unit four- sphere $\Omega_4$.. The functions (in the low/zero temperature phase) are given by
 \beq
 H=(\frac{L}{\r})^3,\;\; f=1-(\frac{R_{kk}}{\r})^3,\;\; h=1,
 \label{lowtemperaturephase}
 \eeq
where $L^3= N_c \pi g_s l_s^3$ and  the
coefficient $R_{kk}$ is a free parameter.
In the high temperature phase (with the coordinate $x_4$ compactified) we have,
 \beq
 H=(\frac{L}{\r})^3,\;\; h=1-(\frac{R_{T}}{\r})^3,\;\; f=1
 \label{hightemperaturephase}
 \eeq
where $R_T$ is a free parameter related to the  temperature of the system,
\be
T=\frac{3}{4\pi} (\frac{L^3}{R_T})^{-1/2}
\ee

Proceeding as described in the previous section  we rotate the background
by applying the U-duality already discussed:  four  T-dualities
in the worldvolume coordinates,
uplift to M-theory, boost, reduce to IIA and T dualize back.
The rotated  background is,

 \bea
 & & ds^2= \mathcal{H}(\r)^{-1/2}
 \Big[ - h(\r) dt^2 + d\vec{x}_{123}^2 +f(\r)dx_4^2  \Big] +
  \mathcal{H}(\r)^{1/2}\Big[\frac{d\r^2}{f(\r)h(\r)}+ \r^2 d\Omega_{4}   \Big],
 \nonumber\\
 & & e^{2\phi[final]}= g_s^2 \mathcal{H}^{-\frac{1}{2}},\nonumber\\
 & & F_{6}= a'(\r) dt\wedge dx_1\wedge....\wedge dx_4\wedge d\r
 \label{D4wittenrotated}
 \eea
 where the functions $\mathcal{H}(\r), a(\r)$ now read
 \bea
 & & \mathcal{H(\r)}= H(\r)\cosh^2\beta+ \frac{1}{g_s}  \sinh 2\beta +(\frac{R_T}{L})^{3} \frac{\sinh^2\beta}{ g_s ^2}\\
 & & a(\r)=\frac{1}{R(\r)^2}\Big[ A(\r)\cosh 2\, \beta +(A(\r)^2+ 1)\cosh\beta\sinh\beta - \frac{f(\r) h(\r) \sinh(2\beta)}{2 g_s^2 H(\r)^2}  \Big]
 \eea
and $R(\r)^2=\frac{\mathcal{H(\r)}}{H(\r)}$. As before, one can check explicitly that the background above
is a solution of the eqs. of motion, but for this we must  set either
$R_T=0$ for the low/zero
temperature phase or  set $R_{kk}=0$, compactifying the $x_4$
direction, in the high
temperature phase.

As explained in the previous section, in spite of having
 started with a decoupled background,
after the rotation we have an asymptotically flat space,
indicating that the rotation has coupled the field theory modes to the
gravitational ones.


\subsection{``Rotation'' in p-q directions}\label{pminusqsection}
We will study now a peculiar situation, where we have Dp branes and we
separate a $q$ manifold from the $p+1$ dimensional world-volume.
The internal manifold need not be specified.
We will impose on our U-duality that:
\begin{itemize}
\item{ after the initial T-dualities, we are in IIA background, so that
this is
upliftable to M-theory}
\item{that the boost generates a NS $H_3$ field, or what is equivalent,
that the IIA configuration after the initial T-dualities contains an
electric
$F_4$.}
\end{itemize}
We will assume (though this
need not be the case) that the internal q-manifold is a torus. In that
case, the solution reads (in string frame as usual)\footnote{we are
dealing here with the Lorentz invariant case here. Working with a
non-extremal factor is straightforward. },
\bea
& & ds^2=H^{-1/2}[-dt^2 + dx_{p-q}^2 + d\sigma_q^2]+ H^{1/2}[d\r^2 +\r^2
d\Omega_{8-p}^2],\nonumber\\
& & F_{p+2}=\partial_\r \hat{A}dt \wedge dx_1\wedge...\wedge dx_{p-q}\wedge
d\sigma_q\wedge d\r,\nonumber\\
& & e^{2\phi[initial]}= e^{2\phi(\infty)}H^{\frac{3-p}{2}}.
\label{entrada1}
\eea
We will proceed as follows: first we will do T dualities in the
$x_{p-q}$
directions, this will leave
us in a IIA configuration, we will lift then to M-theory and boost. We
reduce to IIA and T dualize back, all the details are written in Appendix \ref{details}.
We end up this duality-chain with a final configuration
\bea
& & ds_{II,st}^2= g_{tt}dt^2 + \frac{d\vec{x}_{p-2}^2}{H^{1/3}B^{1/2}}+
B^{1/2}\Big(
H^{-2/3}
d\vec{\sigma}_2^2+
H^{1/3}[d\r^2 +\r^2 d\Omega_{8-p}^2 ]\Big),\nonumber\\
& & F_{p+2}=
\partial_\r \hat{A} (\cosh\beta + a_t \sinh\beta)\wedge
d\r\wedge
d\sigma_1\wedge
d\sigma_2 \wedge dx_1 \wedge....\wedge
dx_{p-2}, \nonumber\\
& & H_3= \sinh\beta \partial_\r \hat{A} d\r \wedge d\sigma_1 \wedge
d\sigma_2,\nonumber\\
& & F_p=\partial_\r(a_t) dt\wedge d\r \wedge dx_1 \wedge....\wedge
dx_{p-2},\nonumber\\
& &  e^{2\phi[final]}= B^{\frac{5-p}{2}}H^{\frac{2-p}{3}}
\label{finaliizzz}
\eea
where we have defined (See Appendix \ref{details} for a detailed derivation),
\bea
& & A= H^{-2/3}[H \sinh^2\beta- \cosh^2\beta],\;\;
B= H^{-2/3}[H \cosh^2\beta- \sinh^2\beta],\nonumber\\
& & C= 2H^{-2/3}\sinh\beta \cosh\beta(1-H),\;\;
g_{tt}=\frac{4AB-C^2}{4\sqrt{B}},\nonumber\\
& & a_t=\frac{C}{2B}=\frac{\sinh\beta \cosh\beta (1-H)}{H\cosh^2\beta-\sinh^2\beta},
\label{caxcax}
\eea
Various aspects are worth emphasizing in this final background. Notice
that:
\begin{itemize}
\item{the metric is dual to a field theory with $SO(1,p-2)$ isometry,
since
\beq
 (4 AB-C^2)H^{1/3}=-4\to g_{tt}=-g_{x_i x_i}
\eeq
}
\item{for $p=5$ the dilaton is invariant under the whole U-duality
procedure, since
\beq
e^{2\phi[initial]}= H^{\frac{3-p}{2}} =e^{2\phi[final]}=
B^{\frac{5-p}{2}}H^{\frac{2-p}{3}}
\eeq
}
\item{for $p=5$ the procedure is generating charge of D3 brane
(represented by the $F_{p=5}$) where the D3 branes are extended in the
$p-q$ directions. This is the reason why the dilaton does not change as the
D3 branes do not
couple to it. The field
$H_3$ is also generated.}
\end{itemize}
In following sections, we will study a particular case of this U-duality,
for a situation in which a set of D5 branes wrap a curved two-manifold (but we
also have a black hole in the metric, breaking the $SO(1,3)$ isometry),
the
results are qualitatively the same.
One should emphasize that there is yet another way of generating NS
three-form fields, that is basically starting from NS five branes in IIA
wrapping a three-cycle, see
\cite{Gaillard:2010gy}
for details.

Also, notice that we have imposed that after the first set of T-dualities, the background is solution
of IIA Supergravity (the conditions for this to happen are discussed in Appendix \ref{details}).
Were this not the case, we present in Appendix \ref{anothersolgen} another possible solution generating
technique with some applications.

Now, we will move to study an interesting application of the U-dualities discussed above.
%
%
%
%

\section{Black Holes in $\cn =1$ SUSY Solutions}\label{rotatingn=1}
As anticipated, in this section we will apply the chain of dualities, lift
and boost proposed in \cite{Maldacena:2009mw}, to generate a new solution
starting from a non-extremal solution in Type IIB.
The interest of the original (`seed') solution is that it was  argued to
be
dual to a field theory with minimal SUSY in four dimensions. The field
theory was studied (at weak coupling) in \cite{Andrews:2006aw} and is
basically $\cn =1$ Super-Yang-Mills plus a sets of (massive) KK vector and chiral
superfields that UV
complete the dynamics. This system was well studied and various string
duals
are known (particular solutions describing the field theory at strong
coupling, with VEV's and
certain operators deforming the Lagrangian). Let us briefly describe the
general form of the
string dual. The `seed' background describes the backreaction of a set of
$N_c$ D5 branes
wrapping a  two-cycle inside the resolved conifold. It consists of a
metric, dilaton $\phi(\r)$ and RR three-form $F_3$ and, in string frame,
it reads (the
coordinates used are
[$t,\vec{x},\r, \theta,\varphi,\tilde{\theta},\tilde{\varphi},\psi$]),
\ba
ds^2 &=& e^{ \phi(\rho)} \Big[dx_{1,3}^2 + e^{2k(\rho)}d\rho^2
+ e^{2 h(\rho)}
(d\theta^2 + \sin^2\theta d\varphi^2) +\nonumber\\
&+&\frac{e^{2 g(\rho)}}{4}
\left((\tilde{\omega}_1+a(\rho)d\theta)^2
+ (\tilde{\omega}_2-a(\rho)\sin\theta d\varphi)^2\right)
 + \frac{e^{2 k(\rho)}}{4}
(\tilde{\omega}_3 + \cos\theta d\varphi)^2\Big], \nonumber\\
F_{(3)} &=&\frac{N_c}{4}\Bigg[-(\tilde{\omega}_1+b(\rho) d\theta)\wedge
(\tilde{\omega}_2-b(\rho) \sin\theta d\varphi)\wedge
(\tilde{\omega}_3 + \cos\theta d\varphi)+\nonumber\\
& & b'd\rho \wedge (-d\theta \wedge \tilde{\omega}_1  +
\sin\theta d\varphi
\wedge
\tilde{\omega}_2) + (1-b(\rho)^2) \sin\theta d\theta\wedge d\varphi \wedge
\tilde{\omega}_3\Bigg].
\label{nonabmetric424}
\ea
where $\tilde\omega_i$ are the left-invariant forms of $SU(2)$
\bea\lab{su2}
&&\tilde{\omega}_1= \cos\psi d\tilde\theta\,+\,\sin\psi\sin\tilde\theta
d\tilde\varphi,\;\;\;\tilde{\omega}_2=-\sin\psi
d\tilde\theta\,+\,\cos\psi\sin\tilde\theta
d\tilde\varphi\,\,,\rc
&&\tilde{\omega}_3=d\psi\,+\,\cos\tilde\theta d\tilde\varphi.
\eea
This supersymmetric system was carefully studied in a  series of papers
\cite{Casero:2006pt},
\cite{HoyosBadajoz:2008fw};
where it was shown that there is a combination of background
functions (basically a `change of basis') that move from a set of coupled
BPS equations to a decoupled one that one can solve-up to one function,
that we will call $P(\r)$. We will not insist much with this formalism
here and refer the interested reader to the original work
\cite{Casero:2006pt},
\cite{HoyosBadajoz:2008fw}\footnote{All this formalism was also applied
to the case in which one also adds fundamental matter to the dual QFT or
sources to the background above. We refer the interested reader
to the review \cite{Nunez:2010sf}.
}.

We will be  more restrictive and for the purposes of this paper, we
will
study solutions where the functions $a(\r)=b(\r)=0$
in the background of eq.(\ref{nonabmetric424}). This is just in order to
make our numerics simpler and illustrate the points we want to make. If we
wanted to construct a black hole solution showing explicitly the
transition between R-symmetry breaking and its restoration, we should then
work
without this restriction. Since in all known solutions the
asymptotics of the functions $a(\r)\sim b(\r)\sim e^{-2\r}$ our asymptotic
results will be qualitatively correct, but near the black hole horizon
there could be important differences.

We will then proceed as follows: first we will propose a background
including a black hole with the restrictions mentioned above
($a(\r)=b(\r)=0$). We will then pass it through the
solution generating
machine. This will produce a new background, now including also
$F_5, H_3$, that we will write explicitly (we have checked that the
equations
of motion before and after the U-duality are the
same). We will be
explicit about the asymptotics of each of the functions near the UV and
near the horizon. This will  produce for us a non-extremal
generalization of the
type IIB string background dual to the `baryonic branch of
KS-system' -certainly with the
restriction that our background does not contain the information of
the breaking of the R-symmetry from $Z_{2N}\to Z_2$.
\subsection{A new `seed' solution}

In this section we consider the background presented in equation (\ref{nonabmetric424})
with the restrictions $a=b=0$ but including the non-extremality factors.
In string frame we have,
\bea
& & ds^2_{IIB,s}= e^{\phi}\Big[-h(\r) dt^2 + dx_1^2 + dx_2^2 + dx_3^2
\Big] +
ds_{6,s}^2,\nonumber\\
& &
ds_6^2=
e^{\phi}\Big[\frac{e^{2k}}{s(\r)}d\r^2+\frac{e^{2k}}{4}(\tilde{\omega}_3 +
\cos\theta d\varphi)^2 +e^{2q}(d\theta^2+\sin^2\theta
d\varphi^2)+\frac{e^{2g}}{4}(d\tilde{\theta}^2
+\sin^2\tilde{\theta}d\tilde{\varphi}^2)   \Big],\nonumber\\
& &
F_{(3)} =\frac{N_c}{4}\Bigg[-\tilde{\omega}_1\wedge
\tilde{\omega}_2 +\sin\theta d\theta \wedge
d\varphi\Bigg]
\wedge (\tilde \omega_3 + \cos\theta d\varphi)=\frac{\alpha' N_c}{4}w_3,
\label{abelianconfiguration}
\eea
where $h(\r), s(\r)$ are the non-extremality functions. In the following,
we will set $s(\r)=h(\r)$ which is simply a choice of parametrization; this implies that (on a particular solution) other
background functions
$\{ q(\r),g(\r), k(\r),\phi(\r) \}$
need not take the same values as in the SUSY
background.

The above Ansatz might be familiar to some readers. It
is worth highlighting a key difference with previous work. Since
our goal is to `rotate' this solution we look for
solutions with stabilized dilaton.  Namely, the typical asymptotic
value of the dilaton inherited from the
D5 (or NS5 brane) solution is linear \cite{Callan:1991dj}, more precisely
\cite{GubserJHEP0109:0172001} \cite{Caceres2009},
\be
e^{\phi} \sim  e^{\rho}\rho^{-1/4}
\ee
We are looking for a dilaton that stabilizes at infinity, that is, which asymptotically behaves as:
\be
e^{\phi} \sim  1 + {\cal O}\left(e^{-8/3 \r}\right).
\ee
Our solution is qualitatively characterized by two parameters: one describing the nonextremality of the solution and the other the speed at which the dilaton gets stabilized.

After applying the solution generating technique, we end up
with a new Type IIB background containing $F_5, H_3$ aside from the fields
originally present in eq.(\ref{abelianconfiguration})
(all details are written in
the Appendix \ref{appendix1}),
\bea
& & ds_{IIB,st}^2= H^{-1/2}\Big[ -h dt^2 + dx_1^2+dx_2^2+dx_3^2  \Big]+
e^{2\phi} H^{1/2}\Big[
e^{2k}\frac{d\r^2}{h(\r)}+\frac{e^{2k}}{4}(\tilde{\omega}_3 +
\cos\theta d\varphi)^2 +  \nonumber\\
& &   +e^{2q}(d\theta^2+\sin^2\theta
d\varphi^2)+\frac{e^{2g}}{4}(d\tilde{\theta}^2
+\sin^2\tilde{\theta}d\tilde{\varphi}^2)  \Big],\nonumber\\
& & H^{1/2}=\cosh\beta
\sqrt{e^{-2\phi}-h(\r)\tanh^2\beta},\;\;\; \phi_{new}=\phi(\r),\nonumber\\
& & F_3 = ( \frac{\alpha' N_c}{4}){\cosh \beta}w_3 ,\;\;
H_3 = -\sqrt{h}( \frac{\alpha' N_c}{4})\sinh \beta \, e^{2
\phi}\,  *_6 \, w_3 \nonumber \\
& & F_5 =  - \tanh \beta ( 1+ *_{10})  \textrm{vol} _4 \wedge
d(\frac{h}{H}).
\label{intermediate}\eea
In the following, we will
focus in the limit $\beta\to\infty$, this is the field theory-limit where
the warp factors vanish at infinity. We will then rescale
\be
\tilde N = N \cosh \beta,\;\;\; x_i \rightarrow
\sqrt{\cosh \beta} \sqrt{\tilde N \alpha'} \, x.
\ee
With all the above the $\beta \rightarrow \infty$
limits are finite and the solution is given by
\bea
& & ds_{IIB,st}^2  = \tilde N \alpha'
\Big[ \tilde{H}^{-1/2}( -h dt^2 + dx_i dx^i  )+
 e^{2\phi} \tilde{H}^{1/2} ds_6^2 \Big],\nonumber\\
& & F_3 = \frac{\alpha' \tilde N }{4 } w_3,\;\;\; H_3 =
-\sqrt{h}\frac{\alpha' \tilde N }{4}\, e^{2 \phi}\,  *_6 \, w_3
\nonumber \\
& & F_5 =   -(\alpha' \tilde N)^2 ( 1+ *_{10})  \textrm{vol} _4 \wedge
d(\frac{h}{\tilde H})
\label{zazaza}
\eea
with
\be
\tilde{H}^{1/2}= \sqrt{e^{-2\phi}-h(\r)}
\label{zsazsa}
\ee
It is worth making a few observations:
\begin{itemize}
\item{We have two new classes of solutions. The equations of motion before and after the
`rotation' are the same, thus a solution of the system  of equations (\ref{abelianconfiguration})
furnishes a solution of the form
(\ref{intermediate})-(\ref{zazaza})}.
\item{The non-extremality factor have made its way into the RR five form
and the warp factor
in eq.(\ref{zsazsa}), but the
dilaton and the three forms are unchanged from the extremal case. Notice
that the factor of $\sqrt{h}$ in the NS field $H_3$ is canceled by the
self dual in six dimensions $*_6 \, w_3$. See Appendix \ref{appendix1}
for full details.}
\item{The black hole in the seed solution will likely have negative
specific heat but its dilaton is stabilized which is a crucial difference with solutions considered previously. We will discuss what is the
behavior of the solution after
the rotation.}
\end{itemize}
In the following section we will explicitly describe the asymptotic
behavior of the solution
as well as its numerical presentation
in the background given by  eq.(\ref{abelianconfiguration}).

\subsection{Asymptotics}
We will proceed to study the asymptotics of the  equations of motion and
solve them numerically to obtain the new non-extremal seed solution (\ref{abelianconfiguration}).

\subsubsection{ UV expansion}
A large $\r$  expansion of the functions
$(k,q,g,h,\phi)$, that solve the equations
of motion up to terms decaying faster than $e^{-4\rho }$ is
\footnote{Note that the UV expansion for $h(\r)=e^{-8 x}$ could
have started with a lower order term
such as
$e^{- 2 \r}$, see Section \ref{choice} for a discussion on the choice of parameters.}
\bea
& & h(\r)= e^{-8 x(\r)}\sim 1+ C_2 e^{-\frac{8}{3}\r},\nonumber\\
& & e^{2q(\r)}\sim \frac{N_c}{4}(2\r-1) +
\frac{c}{4}e^{\frac{4}{3}\r},\;\; e^{2g(\r)}\sim \frac{N_c}{4}(1-2\r) +
\frac{c}{4}e^{\frac{4}{3}\r},\nonumber\\
& & e^{2k(\r)}\sim \frac{2c}{3}e^{\frac{4}{3}\r} -
\frac{N_c^2}{6c}(25-40\r+16\r^2)e^{-\frac{4}{3}\r},\;\;
e^{4\phi-4\phi_0}\sim 1+\frac{3N_c^2}{4c^2}(1-8\r)e^{-\frac{8}{3}\r}.
\label{UVexpansion}
\eea
There are two parameters in the expansion above. The parameter $C_2$
characterizes the non-extremal behavior. The parameter $c$
plays a quite important role. It is one of the integration
constants of the BPS eqs. As an expansion in inverse powers of this
constant
it is possible to write
a solution for the background in eq.(\ref{nonabmetric424})-see for example
\cite{HoyosBadajoz:2008fw}, \cite{Nunez:2008wi}-such that when passed
through this solution generating technique, it results in the Baryonic
Branch dual solution of \cite{Butti:2004pk}.
Physically (and in the background before the rotation), the constant $c$
is describing the coupling of the dynamics on the five-brane to gravity
and ultimately to the full string theory, it takes the configuration
out of its `near brane' limit and can be thought of as indicating the
insertion of an irrelevant operator in the QFT lagrangian.

\subsubsection{Near Horizon Asymptotics}
Near the horizon, $\rho=\rho_h$, we expand the equations of motion in
power series up to order 7.
We have found,
\begin{align}\label{eq:exp-hor}
 e^{-8x(\r)} &= x_1 (\r-\r_h) + x_2 (\r-\r_h)^2+ x_3 (\r-\r_h)^3+ \cdots+ x_8(\r-\r_h)^8\nonumber\\
e^{2 q(\r)} &= q_0 + q_1 (\r-\r_h) + q_2 (\r-\r_h)^2+ \cdots+ q_7 (\r-\r_h)^7\nonumber\\
e^{2 g(\r)} &= g_0 + h_1 (\r-\r_h) + g_2 (\r-\r_h)^2+\cdots + g_7 (\r-\r_h)^7\\
e^{2 k(\r)}& = k_0 + k_1 (\r-\r_h) + k_2 (\r-\r_h)^2+\cdots + k_7 (\r-\r_h)^7\nonumber\\
e^{4 \Phi(\r)}& = f_0 + f_1 (\r-\r_h) + f_2 (\r-\r_h)^2+ \cdots + f_7 (\r-\r_h)^7\nonumber\\
\end{align}
Demanding that the expansions in eq.(\ref{eq:exp-hor}) satisfy the
equations of motion of the system  we determine
$x_2,..., h_1... ,g_1..., k_1..., f_1...$
in terms of $x_1, h_0, g_0, k_0, f_0$
and thus, there are only 5 independent parameters
coming from the expansion at the horizon.
Furthermore, $x_1$ is related to the
non-extremality parameter $\alpha$-see
Appendix \ref{eqsofmotionappendix} for the definitions,
\be
\alpha = -\frac{x_1 \sqrt{f_0} g_0 h_0}{8}
\ee
\subsection{Numerics}
\begin{figure}
\begin{minipage}[t]{.45\textwidth}
 \begin{center}
 \includegraphics[width=7cm]{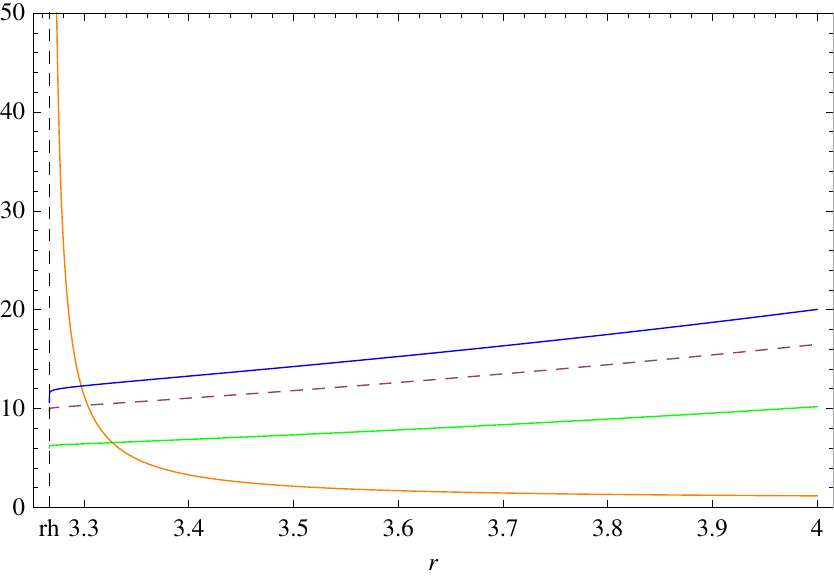}
 \caption{\small{Solution before the rotation for $c=1.97$, $C_2= -7000$. The blue, dashed, green and orange  lines represent  $e^{q}, e^{k}, e^{g}$ and $h^{-1}$ respectively.} }
\label{fig:sol1-97}
\end{center}
\end{minipage}
\begin{minipage}[t]{.45\textwidth}
\begin{center}
 \includegraphics[width=7cm]{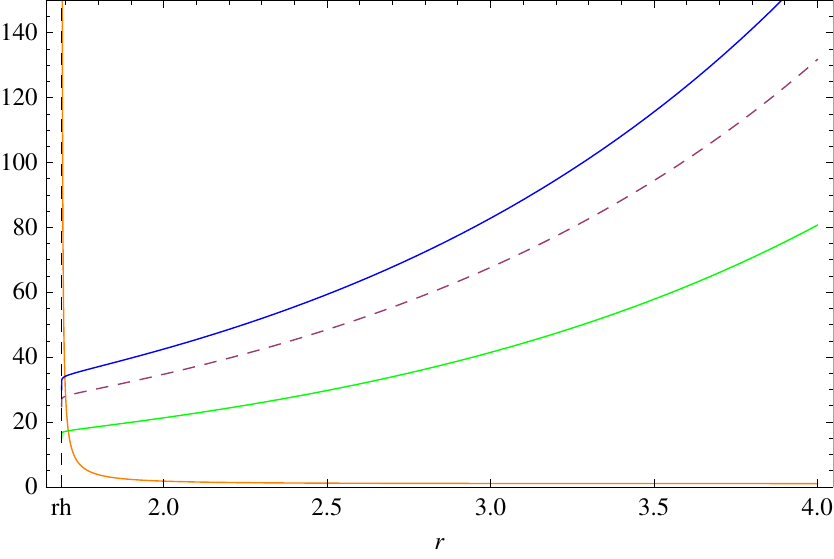}
\caption{\small{Solution before the rotation for $c=125.62$, $C_2= -89$. The blue, dashed, green ad orange  lines represent  $e^q, e^k, e^g$ and $h^{-1}$ respectively. }}
\label{fig:sol125}
\end{center}
\end{minipage}
\end{figure}

\begin{figure}
\begin{minipage}[t]{.45\textwidth}
\begin{center}
 \includegraphics[width=7cm]{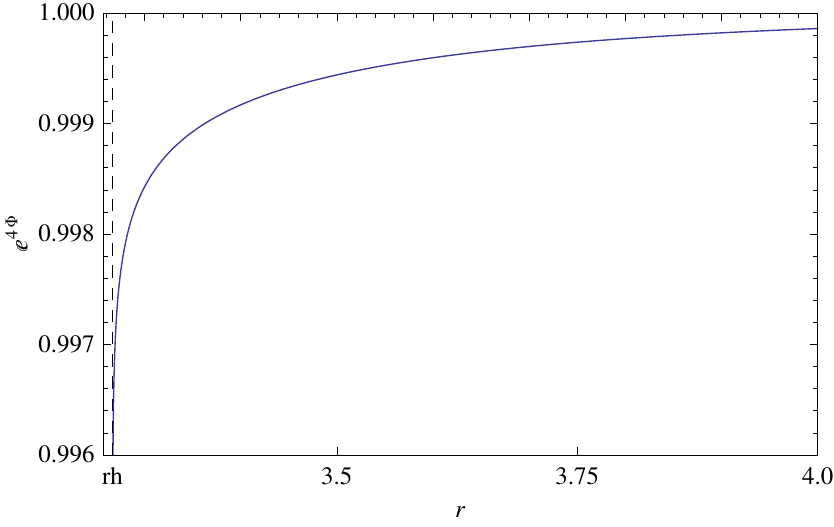}
 \caption{\small{Dilaton, $e^{4\Phi}$ for $c=1.97,  C_2=-7000.$}}
\label{fig:dilc1-97}
\end{center}
\end{minipage}
\begin{minipage}[t]{.45\textwidth}
\begin{center}
 \includegraphics[width=7cm]{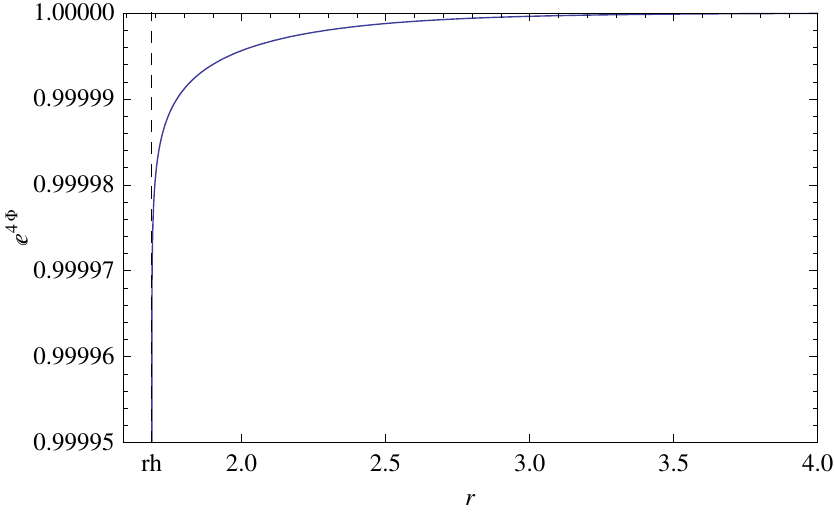}
 \caption{\small{Dilaton, $e^{4\Phi}$ for $c=125.62,  C_2=-89$.}}
\label{fig:dilc125}
\end{center}
\end{minipage}
\end{figure}

The strategy we will follow  is to numerically integrate back from infinity the
equations
of motion in Appendix \ref{eqsofmotionappendix}, using as boundary
conditions the expansion in the UV-eq.(\ref{UVexpansion})-and require that
at the horizon this numerical solution matches
the expansion (\ref{eq:exp-hor})
and its first derivatives -up to seventh order.
Doing this for a given set of values $(c, C_2)$, determines the free
parameters
$(x_1, h_0,
g_0, k_0, f_0)$. Note that not for every pair of $(c, C_2)$ there is a black
hole solution. For some values of $(c, C_2)$ there will not be a horizon and for others there might be
naked singularities outside the horizon. The study of the
this two-parameter family of solutions for the full  $(c, C_2)$
parameter space is an interesting
question that  we will not address here.
We concentrate on finding particular values of $(c, C_2)$
that will produce black hole solutions. Needless to say,
finding a solution before the rotation guarantees that we will have a
solution after the rotation.
Examples of numerical solutions are shown in Figure \ref{fig:sol1-97} and
\ref{fig:sol125}  and the respective
dilatons in Figures \ref{fig:dilc1-97} and \ref{fig:dilc125}.

\subsubsection{Comments about the numerical method}
As mentioned above we integrate back from infinity and match with the expansion at the horizon. In order to do so we define a ``mismatch'' function
evaluated at a point $\r_0$ close to the horizon, $\r_0=\r_h + 0.05$
\begin{align}\label{eq:mismatch}
m(\r_0)=&[g_{sh}(\r_0) -g_{num}(\r_0)]^2 + [g_{sh}'(\r_0) -g'_{num}(\r_0)]^2+[h_{sh}(\r_0) -h_{num}(\r_0)]^2 + \nonumber\\
&[h_{sh}'(\r_0) -h'_{num}(\r_0)]^2
+[k_{sh}(\r_0) -k_{num}(\r_0)]^2 + [k_{sh}'(\r_0) -k'_{num}(\r_0)]^2 +\nonumber\\
&[\Phi_{sh}(\r_0) -\Phi_{num}(\r_0)]^2 + [\Phi_{sh}'(\r_0) -\Phi'_{num}(\r_0)]^2.
\end{align}
The subscript $num$
denotes the numerical solution obtained by
integrating back from infinity and
$sh$  denotes the series expansion  (\ref{eq:exp-hor}) -evaluated at
$\r_0$.
Recall that the expansion at the horizon  depends on five free parameters, $(x_1, h_0,
g_0, k_0, f_0)$. For a given set of $(c, C_2)$ we minimize the mismatch function  to  determine $(x_1, h_0,
g_0, k_0, f_0)$. The parameters at infinity $(c, C_2)$ are adjusted so that
 $m(\r_0) < 5 \times 10^{-5}$. All the numerical procedures where done in Mathematica using  WorkingPrecission 40. 
 We also checked that the constraint $T+U=0$ (see appendix \ref{eqsofmotionappendix}) remains numerically negligible throughout the domain.

\subsection{The rotated solution}
Once we have obtained a numerical solution for the non-extremal background
(\ref{abelianconfiguration}) we can easily generate the rotated solution as summarized in
eq.(\ref{zazaza})
and outlined in the  Appendix \ref{appendix1}. Before presenting the numerical results let us look at
the large $\r$ asymptotics.
Using the UV expansions of the seed solution (\ref{UVexpansion}), one can obtain the black hole metric (\ref{zazaza}) asymptotics {\it after } the rotation,
\bea
& & -g_{tt}\sim \frac{e^{4/3\r}}{A(\r)} -\frac{e^{-4/3\r}}{512 c^4 \sqrt{2}
A(\r)^3}[256c^4 C_2^2 + 27( 1-8\r)^2 + 96 c^2 C_2 ( -1 +
8\r)]+\cdots,\nonumber\\
& & g_{xx}\sim \frac{e^{4/3\r}}{A(\r)} -\frac{e^{-4/3\r}}{512
c^4 \sqrt{2} A(\r)^3}[27( 1- 8\r)^2 ]+\cdots,\nonumber\\
& & g_{\r\r}\sim \frac{2c}{3}A(\r)+ \cdots,\;\;g_{\theta\theta}\sim
\frac{c}{4}A(\r) +\frac{ e^{-4/3\r} }{4}A(\r)(2\r-1) +\cdots,\;\;
\nonumber\\
& & g_{\psi
\psi}\sim \frac{c }{6}A(\r)+ \cdots,\;\;g_{\tilde{\theta}\tilde{\theta}}\sim\frac{c}{4}A(\r) -\frac{e^{-4/3\r}  }{4} A(\r) (2\r-1)+\cdots.
\label{asymptUVrotated}
\eea
Where we have defined the quantity
\beq
c \sqrt{8}A(\r)\equiv\sqrt{24\r -8 C_2 c^2 -3}.
\eeq
Using a variable $u=e^{2 \r/3}$ the metric can be written -to leading order-  as,
\beq
ds^2 = \frac{ u^2}{A(u)} (dx_idx^i ) +\frac{ 3  c A(u) }{2 } (\frac{d u^2}{u^2} +  ds^2_{T^{1 1}}) +   \cdots \mathcal{O}(u^{-2}) .
\eeq
Thus, the  non-extremal solution found after the rotation  has
the asymptotics of the Klebanov-Strassler background
\cite{Klebanov:2000hb}. We point out that the reason we get precisely Klebanov-Tseytlin type of asymptotics lies in our precise arrangements of coefficients in the seed solution. Namely we arranged for a seed solution with no linear dilaton and with a dilaton behavior that, in the field theory limit, eliminates the leading term in the harmonic function given in equation (\ref{zsazsa}).

With our careful choice of asymptotic conditions
for the seed solution of D5 branes we have
 managed to construct black holes in
asymptotically KT backgrounds. Our solution is different from those
obtained in the traditional approach to black holes in
 asymptotically KT backgrounds (see, for example, \cite{KT-blackholes}). One
glaring difference is the form of the RR fluxes.
In particular the presence of the non-extremality parameter in $F_5$ and in the warp factor $H(\r)$ is completely novel.

\begin{figure}
\begin{minipage}[t]{.45\textwidth}
 \begin{center}
 \includegraphics[width=7cm]{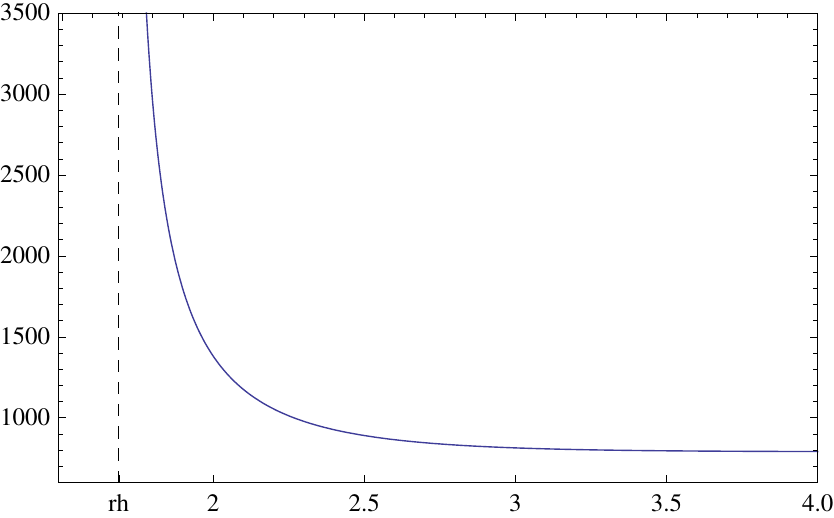}
 \caption{\small{The $g_{r r}$ metric element for a solution after the rotation, $ c=125.62,\  C2= -89$.}
}
\label{fig:grr1after}
\end{center}
\end{minipage}
\begin{minipage}[t]{.45\textwidth}
\begin{center}
 \includegraphics[width=7cm]{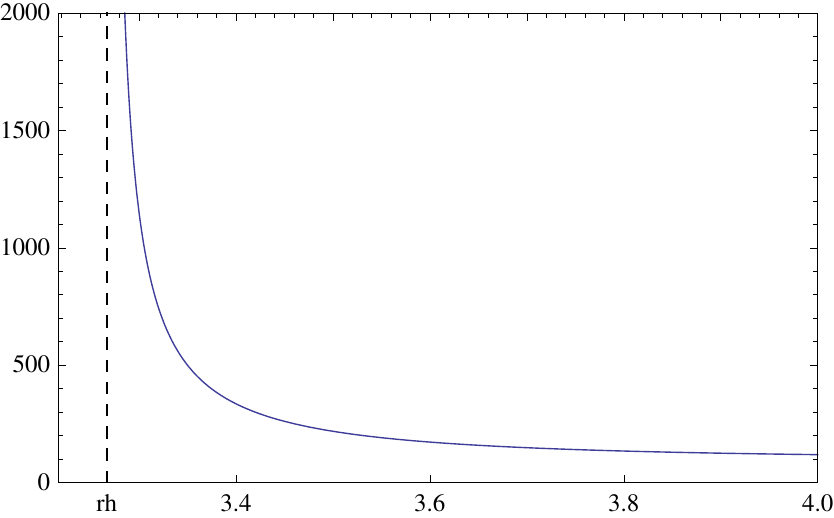}
 \caption{\small{The $g_{r r}$ metric element for a solution after the rotation, $ c=1.97,\  C_2= -7000$.}}
\label{fig:grr2after}
\end{center}
\end{minipage}
\end{figure}

\begin{figure}
\begin{minipage}[t]{.45\textwidth}
 \begin{center}
 \includegraphics[width=7cm]{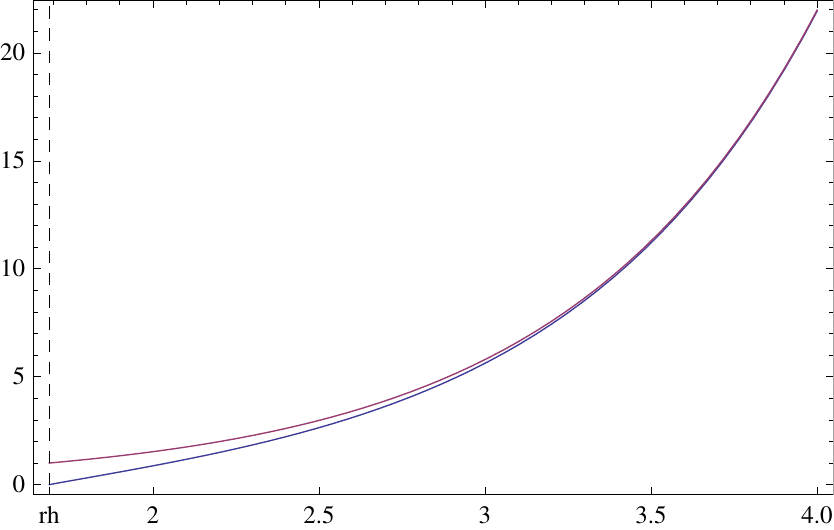}
 \caption{\small{$g_{xx}$ and $g_{tt}$ for a solution after the rotation, $c=125.62,\  C_2= -89$.}}
\label{fig:gxt1after}
\end{center}
\end{minipage}
\begin{minipage}[t]{.45\textwidth}
\begin{center}
 \includegraphics[width=7cm]{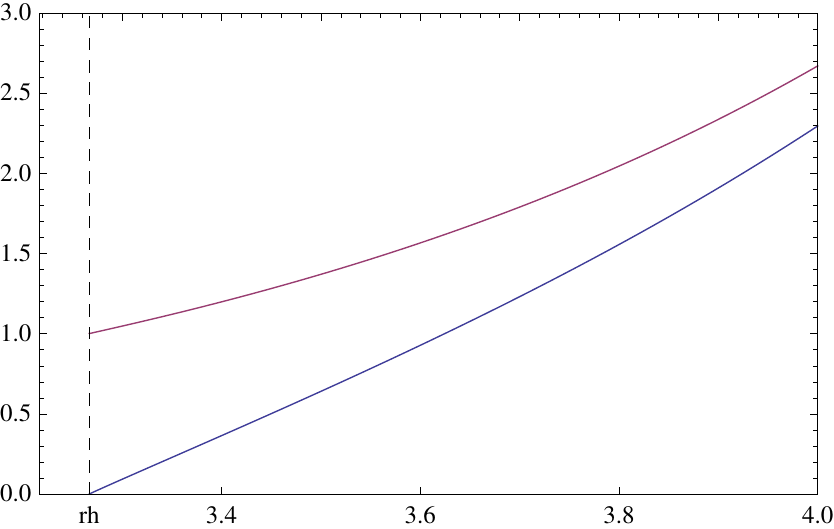}
 \caption{\small{$g_{xx}$ and $g_{tt}$ for a solution after the rotation, $c=1.97,\  C_2= -7000$.}}
\label{fig:gxt2after}
\end{center}
\end{minipage}
\end{figure}

\begin{figure}
\begin{minipage}[t]{.45\textwidth}
 \begin{center}
 \includegraphics[width=7cm]{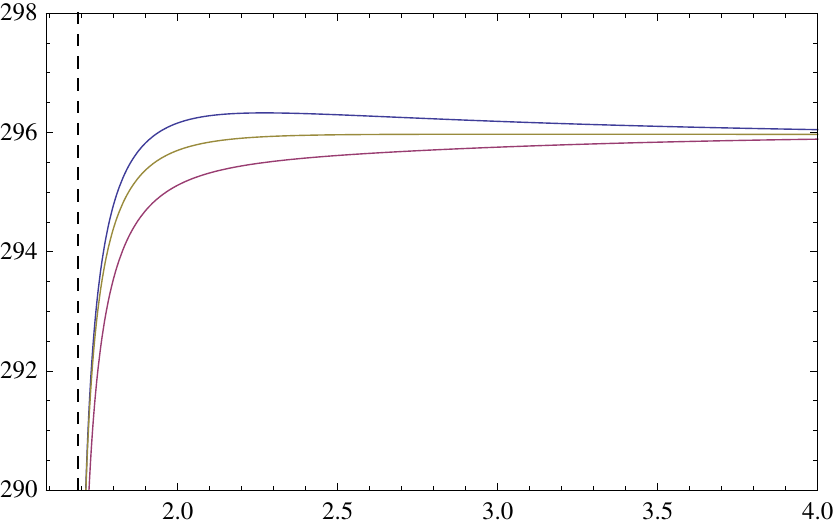}
 \caption{\small{Angular part of the metric after the rotation. We plot $g_{\theta \theta},\  g_{\tilde\theta \tilde\theta}$ and $\frac{3}{2} g_{\psi \psi}$. Solution with $c=125, C_2= -89$.}}
\label{fig:gkh1-T11}
\end{center}
\end{minipage}
\begin{minipage}[t]{.45\textwidth}
\begin{center}
 \includegraphics[width=7cm]{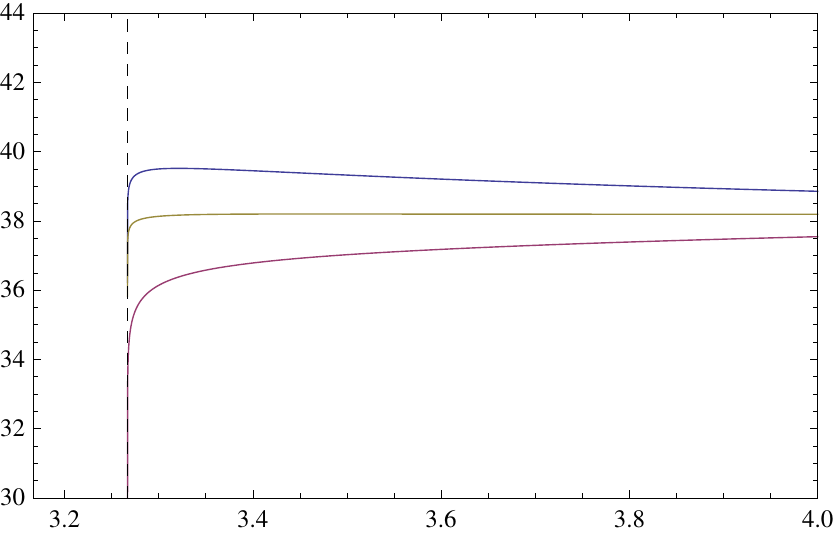}
 \caption{\small{Angular part of the metric after the rotation. We plot $g_{\theta \theta},\  g_{\tilde\theta \tilde\theta}$ and $\frac{3}{2} g_{\psi \psi}$. Solution with $c=1.97, C_2= -7000$.}}
\label{fig:gkh2-T11}
\end{center}
\end{minipage}
\end{figure}

In Figures \ref{fig:grr1after}-\ref{fig:gxt2after} we present the numerical plots of the diffent metric elements  after the rotation for two sets of parameters.
In Figures \ref{fig:gkh1-T11} and \ref{fig:gkh2-T11} we see how the angular part of the solution approaches the $T^{1,1}$ metric for the two
representative solutions.


\subsection{Comments on thermodynamics}\label{analysisinteresting}

Let us clarify the expectations
of the form of the background that arise from field theory considerations.
In the previous section we obtained a
solution largely characterized by two parameters
$(c,C_2)$. After the rotation, the interpretation of $c$ has  been
spelled out explicitly in  \cite{Dymarsky:2005xt}, \cite{Maldacena:2009mw} and
\cite{Gaillard:2010qg}. It is related to the
vev of the baryon operator. The other parameter present in the
solution is $C_2$, its interpretation coincides roughly with the temperature of the solution in the field theory.

We will present and exhaustive analysis of these classes of solutions elsewhere and will limit our attention here to a few main points. The main
deterrent in presenting a final description
of the thermodynamics
is the need to explore a large parameter
space of the solutions as well as the difficulty
in  matching data at asymptotic infinity with
data near the horizon. The analysis is numerically
costly but we emphasize that these are not difficulties of
principle. Our goal in this work is to explicitly show,
through numerical analysis,  the existence of the conjectured solutions.

Let us show that the thermodynamic analysis can, in principle, be done explicitly for the solutions we have. For the temperature, we use the standard approach of looking at the Euclidean section and imposing the absence of conical singularities to determine the temperature of the supergravity backgrounds. Basically, for a metric with Euclidean section of the form
\be
ds^2 = fd\tau^2 + \frac{d\rho^2}{g}+\ldots,
\ee
we find a temperature equal to
\be
T=\frac{1}{4\pi}\sqrt{f'\, g'},
\ee
where prime represents derivative with respect to the radial coordinate and the above expression is evaluated at the horizon. Interestingly, the temperature in both backgrounds, that is, before and after the rotation,  is the same. In terms of the near horizon data we have
\be
T=\frac{1}{4\pi }\, \frac{x_1}{\sqrt{k_0}}.
\ee
It is worth noting that this is the temperature at the
horizon. In the case of asymptotically
flat backgrounds (before the rotation)
we have the option
of computing the temperature as seen by an observer at infinity.
We refer the reader to  \cite{GubserJHEP0109:0172001} for a
detailed discussion of this procedure in the D5 or
NS5 background. The entropy is
computed, as usual, as the area of the horizon. The most complicated
quantity to compute is the free energy which is required to
compute, for example, the specific heat. For the free energy we need to evaluate the
action. In the case of asymptotically
flat solutions for D5/NS5 branes we
can essentially follow  \cite{GubserJHEP0109:0172001}.
A very preliminary analysis seems
to confirm our intuition about the specific heat of such solutions.
Namely, evaluation of the action
suggests that before the rotation
the solutions obtained have negative specific heat.
A precise evaluation of the action after the
rotation is more subtle as naive evaluation
yields divergent terms. It is likely that
a more accurate approach requires some sort
of holographic renormalization or substraction along
the lines of \cite{KT-blackholes}.
Preliminary evaluation suggests
a positive specific heat for the rotated solutions
\footnote{Dario Martelli suggested that along the lines of the first two
papers in \cite{Lunin:2005jy}, some thermodynamical quantities should be
invariant under
T-dualities, shifts of coordinates and boosts. See also
\cite{Horowitz:1993wt}. }.

\section{About decoupling limits}\label{seccion4}
It is worth discussing a bit about decoupling limits. Let us do this in the extremal case, the extension to the cases with non extremal factors follows from what we
write here.
The first point to emphasize is that when gravity modes are coupled to a generic D-brane configuration, the warp factors asymptote to a constant. An example of this
is what happens to the flat Dp branes, with $H=1+(\frac{Q}{\r})^{7-p}$.

We will compare warp factors in the examples studied in the
previous sections.
It is enough to focus on
the expressions of the metric component $g_{tt}$.
In all the examples studied we found
\beq
g_{tt}=\frac{4AB-C^2}{4\sqrt{B}}.
\eeq
Indeed, in the case in which we rotate a flat Dp-brane, we have
eq.(\ref{kajet}) with $A,B,C$ given in eq.(\ref{kaj}).
In the case of the rotation in $p-q$
directions the results for $A,B,C$ are written in eq.(\ref{caxcax})
and in the case of the wrapped D5's are given in
eq.(\ref{xxxx}).
Again, for the point we wish to emphasize, we will focus on
the extremal cases,
when the rotation preserves $SO(1,p)$.
So, in these cases we have\footnote{Gregory Giecold correctly pointed out that
in the non-extremal cases the non-extremality factor $h(\r)$
enters as expected, for example
$g_{tt,D5}=- \frac{h}{\cosh\beta \sqrt{e^{-2\phi}- h\tanh^2\beta}}$. We thank him for this and various other correct comments.
},
\bea
& & g_{tt,(p-q)}=-\frac{1}{\cosh\beta\sqrt{H-\tanh^2\beta}},\;
g_{tt,p}=-
\frac{\sqrt{H}}{\sqrt{H^2(\hat{A}\tanh\beta -1)^2-\tanh^2\beta}},\nonumber\\
& & g_{tt,D5}=- \frac{1}{\cosh\beta \sqrt{e^{-2\phi}- \tanh^2\beta}}
\eea
The case of the rotation in $p$ directions shows that the induced charge of D0 brane
(after the first set of T-dualities) as shown by  eq.(\ref{nnnn}),
enters in the new warp factor via the one-form potential $A_1=\hat{A} dt$.
This makes the decoupling of the final configuration more unclear
and typically, if we start with `decoupled' Dp branes (that is $H\sim \r^{p-7}$)
we will generate a non-decoupled configuration, with an asymptotically
constant warp factor as discussed around eq.(\ref{vvvvxv}).
Of course, we can try to take the decoupling limit in the final configuration.

On the other hand the cases of the wrapped D5's and the rotation in $p-q$
directions (that share the fact that a D0 charge is induced by the boost only)
are such that we {\it must} start with a
non-decoupled configuration in order to
avoid a generated background with a singularity or end of the space
 in the UV (zero warp factor at some finite value of the radial coordinate).
Indeed, if $H= 1+(\frac{Q}{\r})^{7-p}$ we
choose a particular value of the boost parameter $\beta\to\infty$, together with
a particular choice of the value of $e^{\phi(\infty)}=1$ in the D5's case)
to obtain a background decoupled from gravity\footnote{
One of the interesting observations in \cite{Maldacena:2009mw}, is that
in the case of the D5's the solution described in the papers
\cite{HoyosBadajoz:2008fw}, \cite{Nunez:2008wi}
is such that the dilaton asymptotes to a (tunable)
constant and the subleading terms are precisely
those needed to generate the warp factor of Klebanov-Strassler}.






\subsection{Choice of Parameters for ${\cal N}=1$ Black Holes}\label{choice}

Let us discuss in more detail the choice of parameters in the solution. This is a particularly sensitive question from the numerical and from the physical points of view. We note that {\it a priori} there are ten parameters in our solution as corresponds to a system of five second order ordinary differential equations. The constraint lowers this number by one. At this point the simplest route for us is to notice that we are looking for a particular solution  that asymptotes to the extremal supersymmetric one found in \cite{HoyosBadajoz:2008fw},\cite{Gaillard:2010qg} (see, in particular, appendix B there). In that context most of our constants get fixed, as we require the same UV behavior; we use the expressions in that solution to fix six constants. Of the three constants left, one gets fixed by the asymptotic behavior of the non-extremality function, that is, we choose
\be
h(\rho)\sim 1-C_2 e^{-\frac{8}{3}\rho}.
\ee
The leading coefficient guarantees that we will have, after rotation, a solution with KT asymptotics.  The key observation to be made is that, in principle, there is a mode allowed that we have explicitly set to zero. Namely, another solution to the equations of motion in the asymptotic regime could be
\be
h(\rho)\sim 1+C_1 e^{-2 \rho} -C_2 e^{-\frac{8}{3}\rho}+\ldots.
\ee
The above expression should be thought of schematically as the solution to a second order differential equation for $h(\rho)$ which is defined asymptotically by two parameters. This $C_1$ is one of the three constants that we fix and therefore we remain with two constants that we called $(c,C_2)$. The parameter $c$ has the same meaning discussed in \cite{HoyosBadajoz:2008fw},\cite{Gaillard:2010qg} and $C_2$ is related to the temperature.  Thus, the condition that the solution after the rotation has KT asymptotics plus a condition on the non-extremality function determine  the values of the integration constants.

It is worth noting that selecting the right constants in the D3/D5 picture is more cumbersome as there are more parameters in the generic D3/D5 approach to black holes with KT asymptotics. For example, our solution does not allow for a generic mode that will lead to a Freund-Rubin form for $F_5$. As has been repeatedly emphasized, our $F_5$ is related to the non-extremality function and deforms the typical term proportional to the volume $5$-form.  Let us explicitly visualize the difficulty in choosing the asymptotic expansion from the D3/D5 point of view in another example. The choice $C_1=0$  roughly corresponds to a dimension 3 operator in the D3/D5 picture. This can be seen easily by going to the radial coordinate $u=\exp(2\rho/3)$.


\section{Conclusions and Future Directions}
We present here some concluding remarks together with some possible future
projects to complement and refine the material in this paper.

In this work we have elaborated on the U-duality proposed in
\cite{Maldacena:2009mw}. In that case, it was used to connect solutions
describing wrapped D5 branes with the family of
solutions describing the strong coupling dynamics of the baryonic
branch of the Klebanov-Strassler field theory. In our case, we have used
this U-duality to construct a new family of solutions describing
the finite temperature phase of this baryonic branch.
But before doing so, we explored the U-duality (that, following custom,
we called rotation)
in more generality, as a solution generating technique.
We applied it to flat Dp branes in two different circumstances
so that we  appreciated the subtle
differences making the choice of dualities in
\cite{Maldacena:2009mw} so special.

Once we developed this technology, we applied it to rotate the Witten
model for a Yang-Mills dual \cite{Witten:1998zw}.
All this represents new material
and it would be interesting to find an application for it, hopefully in the context
 of duals to Yang-Mills or QCD-like field theories.

In section \ref{rotatingn=1}, we discussed various new things: first
a class of
non-extremal solutions describing five branes wrapping a two-cycle inside
the resolved conifold. The main feature of these non-extremal
backgrounds is that the dilaton is asymptotically stabilized,
which we interpreted as coupling the dual field theory to gravity
modes. Then, we applied the U-duality to this solution, to find
a new background of IIB that describes a non-extremal deformation
of the Klebanov-Tseytlin solution ({\it decoupled} from
gravity or stringy modes). We briefly commented on the
associated thermodynamic properties of this new family of backgrounds.
Finally, we discussed various aspects of
this coupling/decoupling from gravity modes
in section \ref{seccion4}.

Let us now list a set of projects that complement the
material presented here and that may improve the understanding
of the topic:
\begin{itemize}
\item{It is important to sort out the details of the thermodynamics {\it after}
the rotation in the solutions presented in
section \ref{rotatingn=1}. Indeed, as we indicated
in section \ref{analysisinteresting}, this poses a numerical problem, but
certainly does not amount to a conceptual
obstruction. The result for
quantities like the specific heat or the free energy
will be of great interest. Notice that this presents a
novel way to approach the
finite temperature phase of the Klebanov-Strassler field theory.
Comparison with the
solutions of \cite{Bigazzi:2009bk}
would be interesting.}

\item{
More conceptually; it would be nice to better understand the role played by the
D0 branes in this solution generating technique. Indeed, notice the
differences between the examples analyzed in section \ref{warmupz} and
the one in section \ref{psection}, all due to the presence of D0 branes
{\it before} the boost that translates into the different final
results. In the same vein, it would be good to have a
better understanding on the reasons for the equivalence
between the U-duality explained here
and the rotation of the $J_2,\Omega_3$ forms as described in
\cite{Minasian:2009rn}, \cite{Gaillard:2010qg}.

}

\item{
It would be of very much interest to extend our non-extremal solution
to a background of the form in
eq.(\ref{nonabmetric424}) where the restriction $a(\r)=b(\r)=0$
is not applied. Indeed, finding such a non-extremal solution
would, after the application of the U-duality,
produce a solution dual to a non-SUSY KS-like
field theory in the baryonic branch.
It may prove interesting to compare this new
solution with that presented in the papers \cite{Schvellinger:2004am}.
The different nature of the non-SUSY
deformation will certainly reflect in different field theory aspects.
}

\item{It would be nice to use the solution presented
in section \ref{rotatingn=1} as a base to produce a non-SUSY deformation
along the line proposed recently in \cite{Bena:2009xk}.
The authors of \cite{Bena:2009xk} are finding
SUSY breaking backgrounds by adding anti-D3 branes. Their solutions
have singularities (and the
problem is if such a singularity is physically acceptable). Proceeding as
with the solution in this paper may set us on a different branch of solutions.
}
 \item{It would be interesting to find a Physics
application of the new backgrounds of section
\ref{wittensakaisugimotosection}.
}

\item{It would be nice to extend the treatment here to the
case in which the seed background contains
$N_f\sim N_c$ flavors. This would be following the
lead of the papers \cite{Caceres2007},
\cite{Caceres2009}. The subtle point
would be how to rotate such solutions (in other words, how the
sources get represented in M-theory, see \cite{Gaillard:2009kz}.
Comparison with the
solutions of \cite{Bigazzi:2009bk}
would be interesting)}
\item We have presented two numerical solutions as examples of the type of backgrounds obtained. These examples correspond to specific values of the UV parameteers. We did not attempt a study of the full family of solutions in the whole  parameter space. This is an interesting question worth investigating.
\end{itemize}
We hope to address some of these problems in the near future.

\section*{Acknowledgments:} We would like to thank various
colleagues for the input
that allowed us to better understand and present the results of
this paper. We would like to single out Eloy Ay\'on-Beato, Iosif Bena,
Alex Buchel, Pau
Figueras, Gregory
Giecold,
Mariana Gra\~na, Prem Kumar, Dario Martelli,
Ioannis Papadimitriou, Michela
Petrini, Dori Reichmann and C\'esar Terrero-Escalante.
Very special thanks
to Gregory Giecold for a very careful
reading of the manuscript. E.C. thanks
the Theory Group at the University of Texas at Austin for hospitality.

This work was supported in part  by the National Science Foundation under Grant Numbers PHY-0969020 and PHY-0455649,  the US Department of Energy
under grant DE-FG02-95ER40899 and CONACyT grant CB-2008-01-104649.

\appendix

\section{Appendix: U-duality in the case of flat Dp-branes}\label{facil}
\setcounter{equation}{0}
In this appendix we will give
a sketchy derivation of the U-duality chain, in the
case in which we rotate flat Dp-branes.
We will not consider the non-extremal case, this can be done
easily and the result of doing so is included in the first section of the paper.
Here we want just to emphasize some points.

We start from the SUSY configuration in type II supergravity (string frame),
\bea
& & ds^2=H^{-1/2}(-dt^2 + dx_{p}^2)+ H^{1/2} (d\r^2 +\r^2 d\Omega_{8-p}),\nonumber\\
& & e^{2\phi}= H^{\frac{3-p}{2}},\; F_{p+2}= \partial_\r \hat{A} dt\wedge dx_1\wedge
...dx_p\wedge d\r
\eea
\noindent where $\hat{A}= -A$ in the notation of Section \ref{warmupz}. Following standard convention we define the charge of $F_{8-p}$  as positive. The sign of $F_{p+2}$ will then depend on whether we consider Minkowski or Euclidean worldvolume on the $D_p$ brane. We apply T-duality in the $x_p$ directions,
\bea
& & ds^2=- H^{-1/2}dt^2 + H^{1/2} (dx_{p}^2+d\r^2 +\r^2 d\Omega_{8-p}),\nonumber\\
& & e^{2\phi}= H^{\frac{3}{2}},\; F_{2}= \partial_\r \hat{A} dt\wedge d\r
\label{nnnn}\eea
We lift this to M-theory, using $e^{\frac{4\phi}{3}}=H$,
\bea
& & ds_{11}^2= H(dx_{11}+ \hat{A}dt)^2 -H^{-1}dt^2 + dM_9,\;\;
 dM_9= (dx_{p}^2+d\r^2 +\r^2 d\Omega_{8-p})
\eea
we now boost in the $t-x_{11}$ direction with rapidity $\beta$
\beq
dt\to \cosh\beta dt - \sinh\beta dx_{11}, \;\;
dx_{11}\to -\sinh\beta dt + \cosh\beta dx_{11},
\eeq
to get
\bea
& & ds_{11}^2= A dt^2 + B dx_{11}^2 + C dt dx_{11} + dM_9,\nonumber\\
& & A= H^{-1}[H^2(\hat{A}\cosh\beta-\sinh\beta)^2 - \cosh^2\beta]\nonumber\\
& & B= H^{-1}[H^2(\hat{A}\sinh\beta-\cosh\beta)^2 - \sinh^2\beta]\nonumber\\
& & C= -2H^{-1}[H^2 (\hat{A}\cosh\beta-\sinh\beta) (\hat{A}\sinh\beta-\cosh\beta)
- \cosh\beta\sinh\beta]
\label{kaj}
\eea
and we prepare the metric for reduction back to IIA as
\bea
& & ds_{11}^2= B(dx_{11}- a_t dt)^2+ B^{-1/2}\Big[g_{tt}dt^2 + B^{1/2}dM_9   \Big],\nonumber\\
& & a_t=-\frac{C}{2B},\;\;\; g_{tt}=\frac{4AB-C^2}{4\sqrt{B}}
\label{kajet}
\eea
we now write the IIA configuration as,
\bea
& &  ds_{IIA}^2= g_{tt}dt^2 + B^{1/2}(dx_{p}^2+d\r^2 +\r^2 d\Omega_{8-p}),\;
e^{2\phi}= B^{3/2},\; F_2=\partial_\r a_t dt\wedge d\r
\eea
and we T-dualize back in the $x_p$ directions to get
\bea
& & ds_{II}^2= g_{tt}dt^2 +B^{-1/2} dx_p^2+ B^{1/2}(d\r^2 +\r^2 d\Omega_{8-p}),\nonumber\\
& & e^{2\phi}= B^{\frac{3-p}{2}},\;\;\; F_{p+2}=\partial_\r a_t dt\wedge d\r\wedge dx_1 ....\wedge dx_p
\label{dbranekjtfinal}\eea
We see that the $SO(1,p)$ invariance is preserved, as
$g_{tt}\sqrt{B}=-1$.

\section{Appendix: Black Dp branes}\label{Dpblack}
\setcounter{equation}{0}
In this appendix we  gather some useful
facts about black Dp-brane solutions
\cite{Horowitz:1991cd},\cite{Duff:1993ye},\cite{Duff:1994an}.
The metric in string frame is,
\be\label{eq:metric-orig}
ds^2=-\frac{f_{+}(r)}{\sqrt{f_-(r)}}dt^2 +\sqrt{f_-(r)}
\sum_{i=1}^p dx^i dx^i + \frac{f_-(r)^{-\frac{1}{2}-
\frac{5-p}{7-p}}}{f_+(r)}dr^2 + r^2 f_-(r)^{\frac{1}{2}-\frac{5-p}{7-p}}d\Omega^2_{8-p}
\ee
where
\be
f_\pm(r)= 1 - \left(\frac{r_\pm}{r}\right)^{7-p}
\ee
The mass per unit volume $M$  and the R-R charge $N$   are
\be \label{eq:masscharge}
M=\frac{1}{(7-p)(2 \pi)^7 d_p l_P^8} ((8-p) r_+^{7-p} - r_-^{7-p}),\ \ N=c_p( r_+ r_-)^{\frac{7-p}{2}},
\ee
where $l_P$ is the ten dimensional Planck length $l_P= g_s^{1/4} l_s$ and $c_p, \ d_p$ are  numerical factors,
\begin{align}\nonumber
d_p&= 2^{5-p}\pi^{\frac{5-p}{2}} \Gamma\left(\frac{7-p}{2}\right)\nonumber\\ c_p&=\frac{1}{d_p g_s l_s^{7-p}}= \frac{(7-p)Vol_{S^{8-p}}}{(2 \pi l_s)^{7-p} g_s}
\end{align}
The background also has a nontrivial dilaton and a R-R flux,
\begin{align}
e^{-2\Phi} &= g_s^{-2} f_-(r)^{-\frac{p-3}{2}}\nonumber \\
\frac{1}{(2 \pi l_s)^{7-p}}\int_{S^{8-p}} *F_{p+2} &=  N .
\end{align}
 The Einstein frame metric ($g_{E_{\mu\nu}} = \sqrt{g_s e^{-\Phi}} g_{st_{ \mu\nu}}$) has a horizon at $r=r_+$ and, for $p \le 6$, a singularity at $r=r_-$. If $r_+ > r_-$ the singularity is covered by the horizon and the solution is a black hole. In the extremal case, $r_+=r_-$, the space is singular except for $p=3$.

Let us define a new coordinate,
$$ \r^{7-p}= r^{7-p} - r_-^{7-p} $$.
This change of variables transforms (\ref{eq:metric-orig}) to a more familiar form,
\be
ds^2= H(\r)^{-\frac{1}{2}} \left( -h(\r) dt^2 +\sum_{i=1}^p dx^i dx^i\right) +H(\r)^{\frac{1}{2}}\left( \frac{d\r^2}{h(\r)} + \r^2 d\Omega_{8-p}^2\right)
\ee
where
\begin{align}\label{eq:standardHh}
H(\r)&= 1+ \frac{r_-^{7-p}}{\r^{7-p}}\nonumber \\
h(\r)&=1-\frac{r_+ ^{7-p}-r_-^{7-p}}{\r^{7-p}}
\end{align}

Throughout the paper we use the notation,
$$L_p^{7-p} \equiv r_-^{7-p}\ \ \textrm{ and}\ \
R_T^{7-p}\equiv r_+ ^{7-p}-r_-^{7-p}.$$
Note that using (\ref{eq:masscharge})
one can also write
\begin{align}
L_p^{7-p} &= \tilde{Q} \left[\sqrt{1+ \left(\frac{R_T^{7-p}}{2 \tilde{Q}}\right)^2}-\frac{R_T^{7-p}}{2 \tilde{Q}}\right]\nonumber\\
&= \tilde{Q} \ \alpha_p.
\end{align}
where $\tilde{Q}=\frac{N}{c_p}$ and $\alpha_p=(1+ (\frac{R_T^{7-p}}{2 \tilde{Q}})^2)^{1/2}-\frac{R_T^{7-p}}{2 \tilde{Q}}$.

>From (\ref{eq:masscharge}) we can see that the parameter $R_T$ is related to the energy per unit volume above extremality $\epsilon$,
\be
\Delta M= \epsilon = \frac{r_+^{7-p} -r_-^{7-p}}{(2 \pi)^7 d_p l_P^8}
=\frac{{R_T}^{7-p}}{(2 \pi)^7 d_p l_P^8}.
\ee

Note that in the decoupling limit ($\alpha'\rightarrow 0$, energies fixed)
$\epsilon$
remains fixed and  $\alpha_p \rightarrow 1$.

In the extremal case, $r_+ =r_-$, we have $R_T=0$ and $L_p^{7-p}= \tilde{Q}=N d_p g_s l_s^{7-p} $.

\section{Appendix: Some details on the U-duality}\label{details}
\setcounter{equation}{0}
Here we provide details for the derivations of eq.(\ref{finaliizzz}). We
started with
\bea
& & ds^2=H^{-1/2}[-dt^2 + dx_{p-q}^2 + d\sigma_q^2]+ H^{1/2}[d\r^2 +\r^2
d\Omega_{8-p}^2],\nonumber\\
& & F_{p+2}=\partial_\r \hat{A}dt \wedge dx_1\wedge...\wedge dx_{p-q}\wedge
d\sigma_q\wedge d\r,\nonumber\\
& & e^{2\phi[initial]}= e^{2\phi(\infty)}H^{\frac{3-p}{2}}.
\eea
The we perform the U-duality described in the text
below eq.(\ref{entrada1})
\bea
& & ds^2=H^{-1/2}[-dt^2 + d\sigma_q^2]+ H^{1/2}[d\r^2 +\r^2
d\Omega_{8-p}^2 + dx_{p-q}^2],\nonumber\\
& & F_{q+2}=\partial_r \hat{A}dt \wedge
d\sigma_q\wedge d\r,\nonumber\\
& & e^{2\phi}= e^{2\phi(\infty)}H^{\frac{3-q}{2}} .
\label{tdualitieszz}\eea
Now, we need this to be a configuration in  Type IIA (in order to lift
this to M-theory). We will also want to impose that when lifted to
M-theory this will produce a four-form field $G_4$. If
this is the case,
we must have that either $q=2$ or that $q=4$ (the cases of $q=0, q=6$ are
analogous to what we analyzed in the first section). The case $q=2$, on
which we will elaborate upon below, has the peculiarity that the reduction
from eleven dimensions back to IIA will generate a NS $H_3$ field,
proportional to the boost. Let us see this in detail. To begin with, we
will lift to eleven dimensions the configuration in
eq.(\ref{tdualitieszz}).
\bea
& & ds_{11}^2= H^{1/3}dx_{11}^2 + H^{-2/3}[-dt^2 + d\vec{\sigma}_2^2]+
H^{1/3}[d\r^2 +\r^2 d\Omega_{8-p}^2 + d\vec{x}_{p-2}^2],\nonumber\\
& & G_4= \partial_\r \hat{A} dt \wedge d\r\wedge d\sigma_1\wedge d\sigma_2
\eea
and now we boost with rapidity $\beta$
\beq
dt\to \cosh\beta dt - \sinh\beta dx_{11},\;\;\; dx_{11}\to -\sinh\beta dt
+\cosh\beta dx_{11}
\eeq
and we can rewrite the configuration after the boost as,
\bea
& & ds_{11}^2= A dt^2 + B dx_{11}^2 + C dt dx_{11}+ dM_9^2,\nonumber\\
& & G_4= \partial_\r \hat{A} (\cosh\beta dt -\sinh\beta dx_{11})\wedge
d\r\wedge
d\sigma_1\wedge
d\sigma_2, \nonumber\\
& & dM_9^2= H^{-2/3} d\vec{\sigma}_2^2+
H^{1/3}[d\r^2 +\r^2 d\Omega_{8-p}^2 + d\vec{x}_{p-2}^2],\nonumber\\
& & A= H^{-2/3}[H \sinh^2\beta- \cosh^2\beta],\nonumber\\
& & B= H^{-2/3}[H \cosh^2\beta- \sinh^2\beta],\nonumber\\
& & C= 2H^{-2/3}\sinh\beta \cosh\beta(1-H)
\label{boostedzzz}\eea
when we reduce this to IIA we get,
\bea
& & ds_{IIA,st}^2= g_{tt}dt^2 + B^{1/2}\Big(  H^{-2/3} d\vec{\sigma}_2^2+
H^{1/3}[d\r^2 +\r^2 d\Omega_{8-p}^2 + d\vec{x}_{p-2}^2]\Big),\nonumber\\
& & e^{2\phi[A]}= B^{3/2},\;\;\;
g_{tt}=\frac{4AB-C^2}{4\sqrt{B}},\nonumber\\
& & F_4=
\partial_\r \hat{A} (\cosh\beta + a_t \sinh\beta) dt \wedge
d\r\wedge
d\sigma_1\wedge
d\sigma_2, \nonumber\\
& & H_3= \sinh\beta \partial_\r \hat{A} d\r \wedge d\sigma_1 \wedge
d\sigma_2,\nonumber\\
& & F_2=\partial_\r(a_t) dt\wedge d\r,\;\;\; a_t=
\frac{C}{2B}=\frac{\sinh\beta \cosh\beta (1-H)}{H\cosh^2\beta
-\sinh^2\beta},
\label{iiazzz}
\eea
we see that we have generated a NS magnetic field. Finally, we T-dualize
back in the $\vec{x}_{p-1}$ directions, to get
\bea
& & ds_{II,st}^2= g_{tt}dt^2 + \frac{d\vec{x}_{p-2}^2}{H^{1/3}B^{1/2}}+
B^{1/2}\Big(
H^{-2/3}
d\vec{\sigma}_2^2+
H^{1/3}[d\r^2 +\r^2 d\Omega_{8-p}^2 ]\Big),\nonumber\\
& & F_{p+2}=
\partial_\r \hat{A} (\cosh\beta + a_t \sinh\beta)\wedge
d\r\wedge
d\sigma_1\wedge
d\sigma_2 \wedge dx_1 \wedge....\wedge
dx_{p-2}, \nonumber\\
& & H_3= \sinh\beta \partial_\r \hat{A} d\r \wedge d\sigma_1 \wedge
d\sigma_2,\nonumber\\
& & F_p=\partial_\r(a_t) dt\wedge d\r \wedge dx_1 \wedge....\wedge
dx_{p-2},\nonumber\\
& &  e^{2\phi[final]}= B^{\frac{5-p}{2}}H^{\frac{2-p}{3}}.
\label{finaliizzzbis}
\eea
This completes our derivation of eq.(\ref{finaliizzz}).
\section{Appendix: Another solution generating algorithm}\label{anothersolgen}
\setcounter{equation}{0}
We studied two different `solution generating
techniques' and applied them to different backgrounds. All these
`algorithms' were starting with a background solution to the Type II (A
or B) equations of motion, applying a number of T-dualities that would
transform the background into a solution for Type IIA supergravity. Then
we lifted this to eleven dimensions, where a boost was applied (inducing
a one parameter-$\beta$- family of solutions), then reducing to IIA
and T-dualizing back we had our final generated background.

One may wonder what is the algorithm when, after a number of
T-dualities, we end with a background solving the Type IIB
supergravity equations of
motion. In this case, a way of generating a one-parameter family of
solutions may be  S-dualizing. Indeed, given $a,b,c,d$ real numbers
satisfying $ad-bc=1$, we can start with a IIB solution having axion
$\chi$, dilaton $\phi$, RR and NS thre forms $F_3=d C_2$ and $H_3= dB_2$
and five form $F_5= dC_4+ F_3\wedge B_2$ and by S-duality, generate a new
solutions dependent on the three independent real parameters $a,b,c$.
Indeed, the five
form is
left invariant and
the same happens for the Einstein frame metric, while
\bea
& & F_3[new]= b H_3 + a F_3,\;\;\; H_3[new]= d H_3 + c F_3,\nonumber\\
& &  e^{\phi[new]}=\Big((c\chi+d)^2 + c^2 e^{-2\phi}   \Big)
e^{\phi},\;\;\;
\chi[new]=\frac{(a\chi+b)(c\chi+d) + ac e^{-2\phi}}{\Big((c\chi+d)^2 +
c^2 e^{-2\phi}   \Big)}
\eea
Let us focus our attention in the particular set of values $a=d=0,\;
cb=-1$, this is the transformation that interchanges the three forms and
inverts the value of the dilaton (if the axion is initially zero, as we
will assume).

One can then think about applying this transformation expecting to
generate new interesting solutions. As an example, suppose that we start
with solutions
describing D6 branes wrapping a three-cycle inside the deformed conifold.
Those solutions are dual to a (UV completed version of) N=1 SYM
\cite{Atiyah:2000zz}. A particularly interesting solution is given in
section 3.2 of the paper \cite{Brandhuber:2001kq}. The background consist
of metric, dilaton and RR two-form field,
\beq
 ds_{IIA,st}^2= e^{f}[-dt^2+ dx_1^2+ dx_2^2+ dx_3^2 + dM_6^2],\;\;
e^{2\phi}= e^{3f +2 \phi_0},\;\;  F_2= dA_1
\label{andyiia}
\eeq
where all the details (the manifold $M_6$, the functional form of
$f,\phi$ and the one form $A_1$ are given in eqs(57)-(59) of
\cite{Brandhuber:2001kq}). We can the perform three T-dualities in
$x_{1,2,3}$ leading to
\bea
& & ds_{IIB,st}^2= e^{f}[-dt^2 + dM_6^2] + e^{-f}dx_{1,2,3}^2,\;\;
e^{2\phi}= e^{2\phi_0},\nonumber\\
& &  F_5= dA_1\wedge dx_1\wedge dx_2\wedge dx_3
\eea
that is we have generated D3 brane charge and the dilaton is just a
constant. We now move this to Einstein frame, that leaves us-up to a
constant-with the same metric, perform the S-duality metioned above and
T-dualize back. This brings us to the starting point background
(\ref{andyiia})\footnote{One may  perform an S-duality with
parameters $a=0, bc=-1$ and $d$ being free. This will generate a
background in IIB with constant axion-dilaton. The final IIA
configuration is the same as the initial one.}. Something similar occurs
if we start
with  a solution describing D5 branes wrapping a three-cycle inside a G2
holonomy manifold \cite{Maldacena:2001pb},
a configuration dual to N=1 Yang-Mills Chern-Simons in
2+1 dimensions (with its respective UV completion). Notice that this
operation is not the one proposed in \cite{Gaillard:2010gy}, that is the
reason why these authors were able to generate an interesting solution
starting from \cite{Maldacena:2001pb}.

In both these cases described above, we are generating charge of D3 brane,
this is invariant under the S-duality, hence it is expected that the whole
operation brings us back to the initial configuration. A  more
interesting example is to start from the configuration of D4 branes
wrapping a circle with SUSY breaking boundary conditions \cite{Witten:1998zw}
that we discussed before, see Section
\ref{wittensakaisugimotosection}.

In this case we will apply three T-dualities to a IIA configuration. This
will generate charge of D1 brane that the S-duality will interchange with
F1 charge, after T-dualizing back, we will generate a new background (the
$SO(1,3)$ isometry will be spoiled, so we may want to start with the high
Temperature dual, hence having a non-extremal factor $h(\r)$ in
front of $dt^2$ and the function $f(\r)=1$ in front of $dx_4^2$, but let
us keep things general). Let us
see
some details. After
the
T-dualities, the configuration reads
\bea
& & ds_{IIB,st}^2= H^{-1/2}(-hdt^2 + f dx_4^2)+ H^{1/2}(\frac{d\r^2}{h} + \r^2
d\Omega_4^2 + dx_{1,2,3}^2),\nonumber\\
& & e^{2\phi}= g_s^2 H,\;\;\; F_{3}=\partial_\r A dt\wedge dx_4 \wedge d\r
\eea
now, we need to move this to Einstein frame, multiplying the metric by
$e^{-\frac{\phi}{2}}$, perform the S-duality that will generate $H_3$  and
a
dilaton $e^{-2\phi}= g_s^2 H$. Then, T-dualize back in $x_{1,2,3}$.
The final configurations is,
\bea
& & ds_{IIA,st}^2= \frac{c^{1/4}}{g_s H}(-h dt^2 + fdx_4^2) +
\frac{ d\r^2}{h} + \r^2
d\Omega_4^2 + dx_{1,2,3}^2,\nonumber\\
& & H_3=c \partial_\r A dt\wedge dx_4 \wedge d\r,\;\;\; c^4 e^{-2\phi}=
g_s^2 H.
\eea
As a final remark; had we chosen to T-dualize only in the $x_4$ direction
(in the supergravity approximation, this can be done in the high
Temperature phase only), we would have generated D3 branes, hence after
the S-duality and T-duality, we would be back to the initial
configuration.
\section{Appendix: U-duality for the wrapped D5 branes Black Hole}\label{appendix1}
\setcounter{equation}{0}
We will describe in detail the action of the solution generating technique
proposed in \cite{Maldacena:2009mw}, when applied to  the background of
eq.(\ref{abelianconfiguration}). These techniques have been applied in a similar context in \cite{Chen:2010bn}.
Let us consider things in
the string frame.
\bea
& & ds^2_{s}= e^{\phi}\Big[-h(\r) dt^2 + dx_1^2 + dx_2^2 + dx_3^2
\Big] +
ds_{6,s}^2\, \nonumber\\
& &
ds_6^2=
e^{\phi}\Big[\frac{e^{2k}}{s(\r)}d\r^2+\frac{e^{2k}}{4}(\tilde{\omega}_3 +
\cos\theta d\varphi)^2 +e^{2q}(d\theta^2+\sin^2\theta
d\varphi^2)+\frac{e^{2g}}{4}(d\tilde{\theta}^2
+\sin^2\tilde{\theta}d\tilde{\varphi}^2)   \Big],\nonumber\\
& &
F_{(3)} =\frac{N_c}{4}\Bigg[-\tilde{\omega}_1\wedge
\tilde{\omega}_2 +\sin\theta d\theta \wedge
d\varphi\Bigg]
\wedge (\tilde \omega_3 + \cos\theta d\varphi),
\eea
where $h(\r), s(\r)$ are the non-extremality functions
and we will choose $h(\r)=s(\r)$ as we did
in section \ref{rotatingn=1}.
We can proceed to rotate it. We will follow the procedure explained  in
\cite{Maldacena:2009mw}. So, let us start by writing the effect of the
first T-duality in the $x_1$ direction (all the expressions below are in
string
frame)
\bea
& & ds^2_{IIA}=e^{f} \Big[ - h dt^2 + dx_2^2 + dx_3^2  \Big] +
e^{-f}dx_1^2 + ds_6^2
,\nonumber\\
& & e^{2\phi_{A}} =  e^{2\phi-f}, \;\;\; F_{4}=F_3 \wedge dx_1
\eea
The function $f=\phi$ will be kept to avoid confusion with the transformed
dilatons. Now, we perform the T-duality in $x_2$
\bea
& & ds^B_{IIA}=e^{f} \Big[ - h dt^2 +  dx_3^2  \Big] +
e^{-f}(dx_1^2 + dx_2^2)+ ds_6^2
,\nonumber\\
& & e^{2\phi_{B}} =  e^{2\phi-2f}, \;\;\; F_{5}=F_3 \wedge dx_1\wedge dx_2
(1+*_{10})
\eea
T-dualizing in $x_3$, we get
\bea
& & ds^2_{IIA}=e^{f} \Big[ - h dt^2   \Big] +
e^{-f}(dx_1^2 + dx_2^2 +dx_3^2)+ ds_6^2
,\nonumber\\
& & e^{2\phi_{A}} =  e^{2\phi-3f}, \nonumber\\
& & F_{6}=F_3 \wedge dx_1\wedge dx_2 \wedge dx_3\to F_4=
e^{2f}\sqrt{h} *_{6}F_3 \wedge dt
\label{mamamaka}
\eea
Notice that
\beq
*_{6}F_3= \frac{N_c}{\sqrt{h}}\Big[-2 e^{2q-2g}\sin\theta d\theta \wedge
d\varphi +\frac{e^{2g-2q}}{8}\sin\tilde{\theta}d\tilde{\theta}\wedge
d\tilde{\varphi}   \Big]\wedge d\r .
\label{mamamake}
\eeq
The factor of $\sqrt{h}$ in the $F_4$ of eq.(\ref{mamamaka}) is present to
cancel the factor of $\sqrt{h}$ in the denominator of eq.(\ref{mamamake}).
Now, we lift this to M-theory;
\bea
& & ds^2_{11}=  e^{4/3\phi-2f}dx_{11}^2 + e^{f-2/3\phi}\Big[   - h
e^{f}dt^2
+
e^{-f}(dx_1^2 + dx_2^2 +dx_3^2)+ ds_6^2 \Big] ,\nonumber\\
& & G_{4}=\sqrt{h}e^{2f}*_{6}F_3 \wedge dt.
\eea
We boost in the $t-x_{11}$ directions according to,
\beq
dt\to \cosh\beta dt-\sinh\beta dx_{11},\;\;\;\; dx_{11}\to -\sinh\beta dt
+ \cosh\beta dx_{11}
\eeq
and now we rewrite this boosted metric as,
\bea
& & ds^2_{11}=  e^{f-2/3\phi}\Big[ e^{-f}(dx_1^2 + dx_2^2 +dx_3^2)+ds_6^2
\Big] + A dt^2 + B dx_{11}^2 + C dtdx_{11},\nonumber\\
& & G_{4}=\sqrt{h}e^{2f}*_{6}F_3 \Big[\cosh\beta dt-\sinh\beta dx_{11}
\Big]
\label{mtheoryboosted}
\eea
where,
\bea
& & A=  e^{2f-2/3\phi}[\sinh^2\beta e^{2\phi-4f}- h \cosh^2\beta]
,\;\;B=  e^{2f-2/3\phi}[\cosh^2\beta e^{2\phi-4f}- h \sinh^2\beta]
,\nonumber\\
& & C=  -2\cosh\beta \sinh\beta  e^{2f-2/3\phi}[ e^{2\phi-4f}- h
].
\label{xxxx}\eea
Now, we will reduce this to IIA, before doing so and in order to reduce
to IIA, it is useful to rewrite
eq.(\ref{mtheoryboosted}) as,
\bea
& & ds^2_{11}=  B^{-1/2} \Big[  g_{tt}dt^2 + B^{1/2} e^{f-2/3\phi}(
e^{-f}(dx_1^2 + dx_2^2 +dx_3^2)+ds_6^2) \Big]
+
B(dx_{11}+a_t
dt)^2,\nonumber\\
& & G_{4}=\sqrt{h}e^{2f} *_{6}F_3 \Big[(\cosh\beta +a_t \sinh\beta)
dt-\sinh\beta (dx_{11}+a_t dt)
\Big]
\eea
where we have defined
\beq
a_t=\frac{C}{2B},\;\;\;\; g_{tt}=\frac{4AB-C^2}{4\sqrt{B}},\;\;
e^{4/3\phi_{A}}= B.
\eeq
Now, we reduce to IIA, obtaining in string frame,
\bea
& & ds^2_{IIA}=  g_{tt}dt^2 +
\sqrt{B} e^{-2/3\phi}( dx_1^2 + dx_2^2 +dx_3^2)+
\sqrt{B}e^{f-2/3\phi}ds_6^2,\nonumber\\
& & e^{2\phi_{A}} =  B^{3/2}, \nonumber\\
& & F_{4}=  \sqrt{h}e^{2f}*_{6}F_3 \wedge\Big[(\cosh\beta +a_t
\sinh\beta)dt\Big]
,\nonumber\\
& & H_3=  -\sinh\beta \sqrt{h}e^{2f} *_6 F_3,\nonumber\\
& & F_2=a_t' d\r \wedge dt
\eea
Now, we proceed to do the T-dualities back; T-dualizing in the $x_3$
direction we have
\bea
& & ds^2_{IIB}=  g_{tt}dt^2 +
\sqrt{B} e^{-2/3\phi}( dx_1^2 + dx_2^2 )+
\frac{e^{2/3\phi}}{\sqrt{B}}dx_3^2+
\sqrt{B}e^{f-2/3\phi}ds_6^2,\nonumber\\
& & e^{2\phi_{B}} =  B e^{2/3\phi}, \nonumber\\
& & F_{5}=  \sqrt{h}e^{2f}*_{6}F_3 \wedge \Big[(\cosh\beta +a_t
\sinh\beta)dt
\wedge
dx_3 \Big](1+*_{10})
,\nonumber\\
& & H_3=  -\sinh\beta \sqrt{h}e^{2f} *_6 F_3,\nonumber\\
& & F_3=a_t' d\r \wedge dt\wedge dx_3
\eea
now, we T-dualize in $x_2$
\bea
& & ds^2_{IIA}=  g_{tt}dt^2 +
\sqrt{B} e^{-2/3\phi}( dx_1^2  )+
\frac{e^{2/3\phi}}{\sqrt{B}}(dx_3^2+ dx_2^2)+
\sqrt{B}e^{f-2/3\phi}ds_6^2,\nonumber\\
& & e^{2\phi_{A}} =  \sqrt{B} e^{4/3\phi}, \nonumber\\
& & F_{6}=  \sqrt{h}e^{2f} *_{6}F_3 \wedge \Big[(\cosh\beta +a_t
\sinh\beta)dt
\Big]\wedge
dx_3 \wedge dx_2
,\nonumber\\
& & H_3=  -\sinh\beta \sqrt{h}e^{2f} *_6 F_3,\nonumber\\
& & F_4=a_t' d\r \wedge dt\wedge dx_3\wedge dx_2
\eea
finally, we T-dualize in $x_1$
\bea\label{eq:finalconf}
& & ds^2_{IIB}=  g_{tt}dt^2 +
\frac{e^{2/3\phi}}{\sqrt{B}}(dx_3^2+ dx_1^2 + dx_2^2 )
+\sqrt{B}e^{f-2/3\phi}ds_6^2,\nonumber\\
& & e^{2\phi_{B}} =  e^{2\phi}, \nonumber\\
& & F_7=  \sqrt{h}e^{2f} *_{6}F_3 \wedge \Big[(\cosh\beta +a_t
\sinh\beta)dt \Big]
\wedge
dx_3 \wedge dx_2  \wedge dx_1 \quad \quad F_3 = *_{10} F_7
,\nonumber\\
& & H_3=  -\sinh\beta \sqrt{h}e^{2f} *_6 F_3,\nonumber\\
& & F_5=a_t' d\r \wedge dt\wedge dx_3\wedge dx_2\wedge dx_1 (1+*_{10}).
\eea
After using $f=\phi$ and the definitions for $A,B,C,a_t$ this encodes the
result of eq.(\ref{zazaza}).

\section{Appendix: The equations of motion}\label{eqsofmotionappendix}
\setcounter{equation}{0}
In this appendix we will
quote the equations of motion that we are numerically solving. Our goal is to find a black hole of the metrics described in the main text. In
\cite{GubserJHEP0109:0172001}
the authors studied a very general Ansatz for non-extremal deformations of  $NS5$ branes wrapped on $S^2$.
Their Ansatz can be adapted to our case. It reads --in Einstein frame,
\begin{align}\label{eq:metric}
ds^2 &=  -Y_1\,dt^2+Y_2\,d{\rm x}^n d{\rm x}^n + Y_3\, d\rho^2 + Y_4\,(e_1^2 + e_2 ^2) + Y_5\, \left( (w^1) ^2 + (w^2)^2
\right) + Y_6\,(w^3 +A )^2 \,
\end{align}
Inserting this ansatz in the supergravity action we get,
\be\label{eq:L}
L=  \sum\limits_{i, j} G_{i j}(Y) Y_i'Y_j'- U(Y)= T-U
\ee
Using a parametrization to make $G_{ij} $ diagonal and
choosing the appropiate gauge to make contact with our ansatz we have,
\bea
T&=& e^{2 (g + q - 4 x + \Phi)}\frac{1}{8}\left(\frac12 \left(g'{}^2 +q'{}^2 -2\Phi'{}^2\right)+ 2(g'q'-k'x') + (g'+q'+\Phi')(k'-4x'+2\Phi')\right), \nonumber \\
U&=&\frac{1}{256}e^{-2(g + q -\Phi)}
\Big(-16 e^{2(g+q +k)}(e^{2g} + 4 e^{2q})+(e^{4g} +16 e^{4q})(1+ e^{4k})\Big).
\eea
and,
\begin{align}
& Y_1= e^{ \Phi/2}e^{-8x}, \qquad Y_2=e^{ \Phi/2},\qquad Y_3=e^{\Phi/2} e^{8x} e^{2k},\nonumber\\
& Y_4=e^{\Phi/2} e^{2g},  \qquad Y_5=e^{\Phi/2} e^{2q}.\qquad Y_6=\Phi
\end{align}
>From (\ref{eq:L}) we get the second order equations to solve,
\begin{align}\label{eq:eom}
 2 e^{-4 g + 8 x}  + \frac{1}{8} e^{-4 q + 8 x}  - 2 \Phi'( g'  +  q' - 4
x'  +  \Phi')  - \Phi'' & =0 \nonumber \\
-2  x'( g' + q' - 4 x' + \Phi ' ) - x'' & =0 \nonumber \\
 -e^{ 4 k + 8 x} (2 e^{ -4 g} + \frac{1}{8} e^{-4 q} ) + e^{8 x}  ( 2
e^{-4 g} + \frac{1}{8} e^{-4 q})+ 2 k' ( g'+ q' -4 x'+ \Phi')  + k'' & =0
\\
  2 e^{8 x}( e^{4 k- 4 g}  -2 e^{ 2( k - g)} + 1)  + g' (g' + q' - 4 x' +
\Phi') +  g'' & =0\nonumber \\
 e^{8 x}(-\frac{1}{8} e^{ 4 k - 4 q} + e^{ 2 k - 2 q}- \frac{1}{8}e^{-4 q}
)-2 q' (g' + q' - 4 x' + \Phi') -  q'' & =0 \nonumber
\end{align}
and a first order constraint  which is a consequence of reparametrization invariance,
\begin{align}\label{eq:constraint}
 &  \frac{e^{8x}}{2}\Big(\frac{1}{16}(e^{-4 q}+16 e^{-4g})( 1+ e^{4k})
-e^{2k}( e^{-2q} +4 e^{-2g})\Big) + \frac{1}{2} (g'^2 + q'^2 -2 \Phi '^2)
+\nonumber \\ &2 (g'  q'- k'x') +(g'+q'+\Phi ')
   \left(k'-4 x'+2 \Phi
   '\right) =0
\end{align}
The second  equation in (\ref{eq:eom}) can be integrated to yield a first order equation and
(\ref{eq:eom}) becomes
\begin{align}\label{eq:system}
 2 e^{-4 g + 8 x}  + \frac{1}{8} e^{-4 q + 8 x}  - 2 \Phi'( g'  +  q' - 4
x'  +  \Phi')  - \Phi'' & =0 \nonumber \\
x' -\alpha e^{ -2 ( g + q - 4 x + \Phi)} & =0 \nonumber \\
 -e^{ 4 k + 8 x} (2 e^{ -4 g} + \frac{1}{8} e^{-4 q} ) + e^{8 x}  ( 2
e^{-4 g} + \frac{1}{8} e^{-4 q})+ 2 k' ( g'+ q' -4 x'+ \Phi')  + k'' & =0
\\
  e^{8 x}( e^{4 k- 4 g}  -2 e^{ 2( k - g)} + 1 )  + g' (g' + q' - 4 x' +
\Phi') +  g'' & =0\nonumber \\
e^{8 x}(-\frac{1}{8} e^{ 4 k - 4 q} + e^{ 2 k - 2 q}- \frac{1}{8}e^{-4 q}
)-2 q' (g' + q' - 4 x' + \Phi') -  q'' & =0 \nonumber
\end{align}
where $\alpha$ is a non-extremality parameter.

One can verify that the transformation
\be\label{eq:scalingsym}
r\rightarrow e^{2d} r  \qquad  \Phi\rightarrow \Phi +  d \qquad
x\rightarrow x+ \frac{d}{2} \qquad g\rightarrow g\qquad q\rightarrow q
\qquad k\rightarrow k \ee
where $d$ is a constant, leaves (\ref{eq:eom}) and ( \ref{eq:constraint}) invariant and thus, is a symmetry of the equations of motion and constraint.
Another,  obvious, symmetry of (\ref{eq:eom}) and ( \ref{eq:constraint}) is
\be\label{eq:dilsymm}
\Phi \rightarrow \Phi + C \qquad \alpha\rightarrow \alpha e^{2 C}
\ee
with all the other functions unchanged and $C$ a constant. Finally, there is also a translational symmetry
\be\label{eq:transsymm}
r\rightarrow r+ r_0
\ee
that leaves all equations unchanged.
Note that thermodynamic quantities should be invariant under all  these symmetries.


\begin{thebibliography}{99}
\bibitem{Maldacena:1997re}
  J.~M.~Maldacena,
  Adv.\ Theor.\ Math.\ Phys.\  {\bf 2}, 231 (1998)
  [Int.\ J.\ Theor.\ Phys.\  {\bf 38}, 1113 (1999)]
  [arXiv:hep-th/9711200].

\bibitem{Gubser:1998bc}
  S.~S.~Gubser, I.~R.~Klebanov and A.~M.~Polyakov,
  Phys.\ Lett.\  B {\bf 428}, 105 (1998)
  [arXiv:hep-th/9802109].
  E.~Witten,
  Adv.\ Theor.\ Math.\ Phys.\  {\bf 2}, 253 (1998)
  [arXiv:hep-th/9802150].
  N.~Itzhaki, J.~M.~Maldacena, J.~Sonnenschein and S.~Yankielowicz,
  Phys.\ Rev.\  D {\bf 58}, 046004 (1998)
  [arXiv:hep-th/9802042].

\bibitem{Lunin:2005jy}
  O.~Lunin and J.~M.~Maldacena,
  JHEP {\bf 0505}, 033 (2005)
  [arXiv:hep-th/0502086].
  J.~Maldacena, D.~Martelli and Y.~Tachikawa,
  JHEP {\bf 0810}, 072 (2008)
  [arXiv:0807.1100 [hep-th]].
  C.~P.~Herzog, M.~Rangamani and S.~F.~Ross,
  JHEP {\bf 0811}, 080 (2008)
  [arXiv:0807.1099 [hep-th]].
  C.~P.~Herzog, M.~Rangamani and S.~F.~Ross,
  JHEP {\bf 0811}, 080 (2008)
  [arXiv:0807.1099 [hep-th]].

\bibitem{Belinski:2001ph}
  V.~Belinski and E.~Verdaguer,
{\it  Cambridge, UK: Univ. Pr. (2001) 258 p}
  A.~A.~Pomeransky,
  Phys.\ Rev.\  D {\bf 73}, 044004 (2006)
  [arXiv:hep-th/0507250].
  T.~Mishima and H.~Iguchi,
  Phys.\ Rev.\  D {\bf 73}, 044030 (2006)
  [arXiv:hep-th/0504018].
\bibitem{Elvang:2007rd}
  H.~Elvang and P.~Figueras,
  JHEP {\bf 0705}, 050 (2007)
  [arXiv:hep-th/0701035].

















\bibitem{Maldacena:2009mw}
  J.~Maldacena and D.~Martelli,
  JHEP {\bf 1001}, 104 (2010)
  [arXiv:0906.0591 [hep-th]].


\bibitem{Butti:2004pk}
  A.~Butti, M.~Grana, R.~Minasian, M.~Petrini and A.~Zaffaroni,
  JHEP {\bf 0503}, 069 (2005)
  [arXiv:hep-th/0412187].

\bibitem{Klebanov:2000hb}
  I.~R.~Klebanov and M.~J.~Strassler,
  JHEP {\bf 0008}, 052 (2000)
  [arXiv:hep-th/0007191].


\bibitem{Casero:2006pt}
  R.~Casero, C.~Nunez and A.~Paredes,
  Phys.\ Rev.\  D {\bf 73}, 086005 (2006)
  [arXiv:hep-th/0602027].
  R.~Casero, C.~Nunez and A.~Paredes,
  Phys.\ Rev.\  D {\bf 77}, 046003 (2008)
  [arXiv:0709.3421 [hep-th]].




\bibitem{HoyosBadajoz:2008fw}
  C.~Hoyos-Badajoz, C.~Nunez and I.~Papadimitriou,
  Phys.\ Rev.\  D {\bf 78}, 086005 (2008)
  [arXiv:0807.3039 [hep-th]].

\bibitem{Chamseddine:1997nm}
  A.~H.~Chamseddine and M.~S.~Volkov,
  Phys.\ Rev.\ Lett.\  {\bf 79}, 3343 (1997)
  [arXiv:hep-th/9707176].
  J.~M.~Maldacena and C.~Nunez,
  Phys.\ Rev.\ Lett.\  {\bf 86}, 588 (2001)
  [arXiv:hep-th/0008001].

\bibitem{Minasian:2009rn}
  R.~Minasian, M.~Petrini and A.~Zaffaroni,
  JHEP {\bf 1004}, 080 (2010)
  [arXiv:0907.5147 [hep-th]].



\bibitem{Gaillard:2010qg}
  J.~Gaillard, D.~Martelli, C.~Nunez and I.~Papadimitriou,
  Nucl.\ Phys.\  B {\bf 843}, 1 (2011)
  [arXiv:1004.4638 [hep-th]].

\bibitem{Halmagyi:2010st}
  N.~Halmagyi,
  arXiv:1003.2121 [hep-th].

\bibitem{Dymarsky:2005xt}
  A.~Dymarsky, I.~R.~Klebanov and N.~Seiberg,
  JHEP {\bf 0601}, 155 (2006)
  [arXiv:hep-th/0511254].


\bibitem{Andrews:2006aw}
  R.~P.~Andrews and N.~Dorey,
  Nucl.\ Phys.\  B {\bf 751}, 304 (2006)
  [arXiv:hep-th/0601098].
  R.~P.~Andrews and N.~Dorey,
  Phys.\ Lett.\  B {\bf 631}, 74 (2005)
  [arXiv:hep-th/0505107].









\bibitem{Witten:1998zw}
  E.~Witten,
  Adv.\ Theor.\ Math.\ Phys.\  {\bf 2}, 505 (1998)
  [arXiv:hep-th/9803131].



\bibitem{Sakai:2004cn}
  T.~Sakai and S.~Sugimoto,
  Prog.\ Theor.\ Phys.\  {\bf 113}, 843 (2005)
  [arXiv:hep-th/0412141].

\bibitem{Aharony:2006da}
  O.~Aharony, J.~Sonnenschein and S.~Yankielowicz,
  Annals Phys.\  {\bf 322}, 1420 (2007)
  [arXiv:hep-th/0604161].








\bibitem{Gaillard:2010gy}
  J.~Gaillard and D.~Martelli,
  arXiv:1008.0640 [hep-th].










\bibitem{Nunez:2010sf}
  C.~Nunez, A.~Paredes and A.~V.~Ramallo,
  Adv.\ High Energy Phys.\  {\bf 2010}, 196714 (2010)
  [arXiv:1002.1088 [hep-th]].



\bibitem{Nunez:2008wi}
  C.~Nunez, I.~Papadimitriou and M.~Piai,
  Int.\ J.\ Mod.\ Phys.\  A {\bf 25}, 2837 (2010)
  [arXiv:0812.3655 [hep-th]].





\bibitem{GubserJHEP0109:0172001}
S.~S. Gubser, A.~A. Tseytlin, and M.~S. Volkov,
 {\em JHEP} {\bf
  0109:017,2001} (JHEP 0109:017,2001).

\bibitem{Caceres2007}
E.~Caceres, R.~Flauger, M.~Ihl, and T.~Wrase,
 {\em JHEP} {\bf  0803:020,2008} (Nov., 2007).

\bibitem{Caceres2009}
E.~Caceres, R.~Flauger, and T.~Wrase,
  {{ arXiv:0908.4483}}.


\bibitem{Horowitz:1991cd}
  G.~T.~Horowitz, A.~Strominger,
  Nucl.\ Phys.\  {\bf B360}, 197-209 (1991).


\bibitem{Duff:1993ye}
  M.~J.~Duff, J.~X.~Lu,
  Nucl.\ Phys.\  {\bf B416}, 301-334 (1994).
  [hep-th/9306052].
\bibitem{Duff:1994an}
  M.~J.~Duff, R.~R.~Khuri, J.~X.~Lu,
  Phys.\ Rept.\  {\bf 259}, 213-326 (1995).
  [hep-th/9412184].



\bibitem{Atiyah:2000zz}
  M.~Atiyah, J.~M.~Maldacena and C.~Vafa,
  J.\ Math.\ Phys.\  {\bf 42} (2001) 3209
  [arXiv:hep-th/0011256].
  C.~Vafa,
  J.\ Math.\ Phys.\  {\bf 42}, 2798 (2001)
  [arXiv:hep-th/0008142].

  J.~D.~Edelstein and C.~Nunez,
  JHEP {\bf 0104}, 028 (2001)
  [arXiv:hep-th/0103167].


\bibitem{Brandhuber:2001kq}
  A.~Brandhuber,
  Nucl.\ Phys.\  B {\bf 629}, 393 (2002)
  [arXiv:hep-th/0112113].

\bibitem{Maldacena:2001pb}
  J.~M.~Maldacena and H.~S.~Nastase,
  JHEP {\bf 0109}, 024 (2001)
  [arXiv:hep-th/0105049].

\bibitem{Callan:1991dj}
  C.~G.~.~Callan, J.~A.~Harvey and A.~Strominger,
  Nucl.\ Phys.\  B {\bf 359} (1991) 611.



\bibitem{KT-blackholes}
  A.~Buchel,
  Nucl.\ Phys.\  B {\bf 600} (2001) 219
  [arXiv:hep-th/0011146].
  A.~Buchel, C.~P.~Herzog, I.~R.~Klebanov, L.~A.~Pando Zayas and A.~A.~Tseytlin,
  JHEP {\bf 0104} (2001) 033
  [arXiv:hep-th/0102105].
  S.~S.~Gubser, C.~P.~Herzog, I.~R.~Klebanov and A.~A.~Tseytlin,
  JHEP {\bf 0105} (2001) 028
  [arXiv:hep-th/0102172].
  L.~A.~Pando Zayas and C.~A.~Terrero-Escalante,
  JHEP {\bf 0609} (2006) 051
  [arXiv:hep-th/0605170].
  O.~Aharony, A.~Buchel and P.~Kerner,
  Phys.\ Rev.\  D {\bf 76} (2007) 086005
  [arXiv:0706.1768 [hep-th]].
  M.~Mahato, L.~A.~Pando~Zayas and C.~A.~Terrero-Escalante,
  JHEP {\bf 0709} (2007) 083
  [arXiv:0707.2737 [hep-th]].
  A.~Buchel,
  arXiv:1012.2404 [hep-th].

\bibitem{Horowitz:1993wt}
  G.~T.~Horowitz and D.~L.~Welch,
  Phys.\ Rev.\  D {\bf 49}, 590 (1994)
  [arXiv:hep-th/9308077].

\bibitem{Schvellinger:2004am}
  M.~Schvellinger,
  JHEP {\bf 0409}, 057 (2004)
  [arXiv:hep-th/0407152].
  P.~McGuirk, G.~Shiu and Y.~Sumitomo,
  Nucl.\ Phys.\  B {\bf 842}, 383 (2010)
  [arXiv:0910.4581 [hep-th]].

\bibitem{Bena:2009xk}
  I.~Bena, M.~Grana and N.~Halmagyi,
  JHEP {\bf 1009}, 087 (2010)
  [arXiv:0912.3519 [hep-th]].
  I.~Bena, G.~Giecold and N.~Halmagyi,
  arXiv:1011.2195 [hep-th].




\bibitem{Gaillard:2009kz}
  J.~Gaillard and J.~Schmude,
  JHEP {\bf 1002}, 032 (2010)
  [arXiv:0908.0305 [hep-th]].


\bibitem{Bigazzi:2009bk}
  F.~Bigazzi, A.~L.~Cotrone, J.~Mas, A.~Paredes, A.~V.~Ramallo and J.~Tarrio,
  JHEP {\bf 0911}, 117 (2009)
  [arXiv:0909.2865 [hep-th]].
  F.~Bigazzi, A.~L.~Cotrone and J.~Tarrio,
  JHEP {\bf 1002}, 083 (2010)
  [arXiv:0912.3256 [hep-th]].
  F.~Bigazzi and A.~L.~Cotrone,
  JHEP {\bf 1008}, 128 (2010)
  [arXiv:1006.4634 [hep-ph]].
  M.~Mia, K.~Dasgupta, C.~Gale and S.~Jeon,
  Nucl.\ Phys.\  B {\bf 839} (2010) 187
  [arXiv:0902.1540 [hep-th]].
  F.~Bigazzi, A.~L.~Cotrone, J.~Mas, D.~Mayerson and J.~Tarrio,
  arXiv:1101.3560 [hep-th].


\bibitem{Chen:2010bn}
 F.~Chen, K.~Dasgupta, P.~Franche, S.~Katz and R.~Tatar,
 arXiv:1007.5316 [hep-th].



%
%
%
%
%
%
%
%
%
%
%
%
%
%
%
%
%
%
%
%
%
%
%
 \end{thebibliography}
\end{document}